\documentclass[useAMS,usegraphicx]{mn2e}
\usepackage{subfig}

\usepackage{aas_macros}
\usepackage{rotating}

\def\herschel{{\it Herschel}}

\def\lir{$L_{\rm IR}$}

\def\lsun{$L_{\odot}$}

\def\l1.4{L$_{\rm 1.4GHz}$}
\def\s1.4{$S_{\rm 1.4GHz}$}

\def\td{$T_{\rm D}$}
\def\gs{\mathrel{\raise0.35ex\hbox{$\scriptstyle >$}\kern-0.6em
\lower0.40ex\hbox{{$\scriptstyle \sim$}}}}
\def\ls{\mathrel{\raise0.35ex\hbox{$\scriptstyle <$}\kern-0.6em
\lower0.40ex\hbox{{$\scriptstyle \sim$}}}}

\title[SPIRE-mm Photometric Redshifts]
{The {\it Herschel} Multi-Tiered Extragalactic Survey: SPIRE-mm Photometric Redshifts}
\author[I.G.~Roseboom et al.]
{\parbox{\textwidth}{\raggedright I.G.~Roseboom,$^{1,2}$\thanks{E-mail: \texttt{igr@roe.ac.uk}}
R.J.~Ivison,$^{3,2}$
T.R.~Greve,$^{4}$
A.~Amblard,$^{5}$
V.~Arumugam,$^{3}$
R.~Auld,$^{6}$
H.~Aussel,$^{7}$
M.~Bethermin$^{7}$
A.~Blain,$^{8}$
J.~Bock,$^{8,9}$
A.~Boselli,$^{10}$
D.~Brisbin,$^{11}$
V.~Buat,$^{10}$
D.~Burgarella,$^{10}$
N.~Castro-Rodr{\'\i}guez,$^{12,13}$
A.~Cava,$^{12,13}$
P.~Chanial,$^{7}$
E.~Chapin,$^{14}$
S.~Chapman,$^{15}$
D.L.~Clements,$^{16}$
A.~Conley,$^{17}$
L.~Conversi,$^{18}$
A.~Cooray,$^{5,8}$
C.D.~Dowell,$^{8,9}$
J.S.~Dunlop,$^{2}$
E.~Dwek,$^{19}$
S.~Eales,$^{6}$
D.~Elbaz,$^{7}$
D.~Farrah,$^{1}$
A.~Franceschini,$^{20}$
J.~Glenn,$^{21}$
M.~Griffin,$^{6}$
M.~Halpern,$^{14}$
E.~Hatziminaoglou,$^{22}$
E.~Ibar,$^{3}$
K.~Isaak,$^{6}$
G.~Lagache,$^{23}$
L.~Levenson,$^{8,9}$
N.~Lu,$^{8,24}$
S.~Madden,$^{7}$
B.~Maffei,$^{25}$
G.~Mainetti,$^{20}$
L.~Marchetti,$^{20}$
G.~Marsden,$^{14}$
G.~Morrison,$^{26}$
A.M.J.~Mortier,$^{16}$
H.T.~Nguyen,$^{9,8}$
B.~O'Halloran,$^{16}$
S.J.~Oliver,$^{1}$
A.~Omont,$^{27}$
M.J.~Page,$^{28}$
P.~Panuzzo,$^{7}$
A.~Papageorgiou,$^{6}$
C.P.~Pearson,$^{29,30}$
I.~P{\'e}rez-Fournon,$^{12,13}$
M.~Pohlen,$^{6}$
J.I.~Rawlings,$^{28}$
G.~Raymond,$^{6}$
D.~Rigopoulou$^{29,31}$
D.~Rizzo,$^{16}$
G.~Rodighiero,$^{20}$
M.~Rowan-Robinson,$^{16}$
B.~Schulz,$^{8,24}$
Douglas~Scott,$^{14}$
N.~Seymour,$^{28}$
D.L.~Shupe,$^{8,24}$
A.J.~Smith,$^{1}$
J.A.~Stevens,$^{32}$
M.~Symeonidis,$^{28}$
M.~Trichas,$^{33}$
K.E.~Tugwell,$^{28}$
M.~Vaccari,$^{20}$
I.~Valtchanov,$^{18}$
J.D.~Vieira,$^{8}$
M.P.~Viero,$^{8}$
L.~Vigroux,$^{27}$
J.~Wardlow,$^{5}$
L.~Wang,$^{1}$
G.~Wright,$^{3}$
C.K.~Xu$^{8,24}$ and
M.~Zemcov$^{8,9}$}\vspace{0.4cm}\\
\parbox{\textwidth}{\raggedright $^{1}$Astronomy Centre, Dept. of Physics \& Astronomy, University of Sussex, Brighton BN1 9QH, UK\\
$^{2}$Institute for Astronomy, University of Edinburgh, Royal Observatory, Blackford Hill, Edinburgh EH9 3HJ, UK\\
$^{3}$UK Astronomy Technology Centre, Royal Observatory, Blackford Hill, Edinburgh EH9 3HJ, UK\\
$^{4}$Dark Cosmology Centre, Niels Bohr Institute, University of Copenhagen, Juliane Maries Vej 30, 2100 Copenhagen, Denmark\\
$^{5}$Dept. of Physics \& Astronomy, University of California, Irvine, CA 92697, USA\\
$^{6}$Cardiff School of Physics and Astronomy, Cardiff University, Queens Buildings, The Parade, Cardiff CF24 3AA, UK\\
$^{7}$Laboratoire AIM-Paris-Saclay, CEA/DSM/Irfu - CNRS - Universit\'e Paris Diderot, CE-Saclay, pt courrier 131, F-91191 Gif-sur-Yvette, France\\
$^{8}$California Institute of Technology, 1200 E. California Blvd., Pasadena, CA 91125, USA\\
$^{9}$Jet Propulsion Laboratory, 4800 Oak Grove Drive, Pasadena, CA 91109, USA\\
$^{10}$Laboratoire d'Astrophysique de Marseille, OAMP, Universit\'e Aix-marseille, CNRS, 38 rue Fr\'ed\'eric Joliot-Curie, 13388 Marseille cedex 13, France\\
$^{11}$Space Science Building, Cornell University, Ithaca, NY, 14853-6801, USA\\
$^{12}$Instituto de Astrof{\'\i}sica de Canarias (IAC), E-38200 La Laguna, Tenerife, Spain\\
$^{13}$Departamento de Astrof{\'\i}sica, Universidad de La Laguna (ULL), E-38205 La Laguna, Tenerife, Spain\\
$^{14}$Department of Physics \& Astronomy, University of British Columbia, 6224 Agricultural Road, Vancouver, BC V6T~1Z1, Canada\\
$^{15}$Institute of Astronomy, University of Cambridge, Madingley Road, Cambridge CB3 0HA, UK\\
$^{16}$Astrophysics Group, Imperial College London, Blackett Laboratory, Prince Consort Road, London SW7 2AZ, UK\\
$^{17}$Center for Astrophysics and Space Astronomy, 593 UCB, Boulder, Co 80309-0593, USA\\
$^{18}$Herschel Science Centre, European Space Astronomy Centre, Villanueva de la Ca\~nada, 28691 Madrid, Spain\\
$^{19}$Observational  Cosmology Lab, Code 665, NASA Goddard Space Flight  Center, Greenbelt, MD 20771, USA\\
$^{20}$Dipartimento di Astronomia, Universit\`{a} di Padova, vicolo Osservatorio, 3, 35122 Padova, Italy\\
$^{21}$Dept. of Astrophysical and Planetary Sciences, CASA 389-UCB, University of Colorado, Boulder, CO 80309, USA\\
$^{22}$ESO, Karl-Schwarzschild-Str. 2, 85748 Garching bei M\"unchen, Germany\\
$^{23}$Institut d'Astrophysique Spatiale (IAS), b\^atiment 121, Universit\'e Paris-Sud 11 and CNRS (UMR 8617), 91405 Orsay, France\\
$^{24}$Infrared Processing and Analysis Center, MS 100-22, California Institute of Technology, JPL, Pasadena, CA 91125, USA\\
$^{25}$School of Physics and Astronomy, The University of Manchester, Alan Turing Building, Oxford Road, Manchester M13 9PL, UK\\
$^{26}$Institute for Astronomy, University of Hawaii, Manoa, HI 96822, USA ; Canada-France-Hawaii Telescope Corp., Kamuela, HI 96743, USA\\
$^{27}$Institut d'Astrophysique de Paris, UMR 7095, CNRS, UPMC Univ. Paris 06, 98bis boulevard Arago, F-75014 Paris, France\\
$^{28}$Mullard Space Science Laboratory, University College London, Holmbury St. Mary, Dorking, Surrey RH5 6NT, UK\\
$^{29}$Space Science \& Technology Department, Rutherford Appleton Laboratory, Chilton, Didcot, Oxfordshire OX11 0QX, UK\\
$^{30}$Institute for Space Imaging Science, University of Lethbridge, Lethbridge, Alberta, T1K 3M4, Canada\\
$^{31}$Astrophysics, University of Oxford, Keble Road, Oxford OX1 3RH, UK\\
$^{32}$Centre for Astrophysics Research, University of Hertfordshire, College Lane, Hatfield, Hertfordshire AL10 9AB, UK\\
$^{33}$Harvard-Smithsonian Center for Astrophysics, 60 Garden Street, Cambridge, MA 02138, USA}}

\begin{document}

\date{\today}

\pagerange{\pageref{firstpage}--\pageref{lastpage}} \pubyear{2011}

\maketitle

\label{firstpage}

\begin{abstract}
We investigate the potential of submm--mm and submm--mm--radio photometric redshifts using a sample of mm-selected sources as
seen at 250, 350 and 500$\,\mu$m by the SPIRE instrument on \herschel. 
From a sample of 63 previously identified mm-sources with reliable radio identifications in the GOODS-N and Lockman Hole North fields 46 (73 per cent) are found to have detections in at least one SPIRE band. We explore the observed submm/mm colour evolution with redshift, finding that the colours of mm-sources are adequately described by a modified blackbody with constant optical depth $\tau=(\nu/\nu_0)^{\beta}$ where $\beta=+1.8$ and $\nu_0=c/100\,\mu$m. We find a tight correlation between dust temperature and IR luminosity. Using a single model of the dust temperature and IR luminosity relation we derive photometric redshift estimates for the 46 SPIRE detected mm-sources. Testing against the 22 sources with known spectroscopic, or good quality optical/near-IR photometric, redshifts we find submm/mm photometric redshifts offer a redshift accuracy of  $|\Delta z|/(1+z)=0.16 \,(<|\Delta z|>=0.51)$. Including constraints from the radio-far IR correlation the accuracy is improved to $|\Delta z|/(1+z)=0.15 \,(<|\Delta z|>=0.45)$. We estimate the redshift distribution of mm-selected sources finding a significant excess at $z>3$ when compared to $\sim850\,\mu$m selected samples.

\end{abstract}


\section{Introduction}

As is the case in the local Universe, star formation at high redshift
takes place in regions rich in molecular gas and dust. Franceschini et al.\ (1991) and Blain \&
Longair (1993) thus anticipated that the distant galaxies hosting the
majority of this activity would be discovered in the submm/mm
waveband, which benefits from a strong, negative $K$-correction for the rest-frame
far-infrared (-IR) portion of their spectral energy distributions
(SEDs). Observing at 450 and 850$\,\mu$m the SCUBA submm camera on the JCMT (Holland et al.\ 1999) made it possible
to detect the first examples of this galaxy population and represented
a fundamental turning point in our understanding of galaxy formation
in the distant Universe (Smail, Ivison \& Blain 1997; Barger et al.\
1998; Hughes et al.\ 1998; Eales et al.\ 1999; Dey et al.\ 1999).

Early work on the submm galaxy population was often hindered by small numbers of sources, and significant difficulties in finding multiwavelength counterparts given the relatively low angular resolution ({\sc FWHM}$>10$ arcsec) of ground-based submm/mm facilities. This situation has been much improved by a large increase in the number of submm/mm facilities, larger scale submm surveys (e.g. SHADES, Mortier et al.\ 2005; LESS, Wei\ss\ et al.\ 2009), improvements in the quality of overlapping multiwavelength data (in particular radio and mid-IR imaging), as well as advances in cross-matching algorithms (Ivison et al. 2007; Chapin et al.\ 2009; Roseboom et al.\ 2009; Chapin et al.\ 2011). As a result, our understanding of these galaxies has accelerated in recent years (e.g.\ Chapman et al.\ 2005; Greve et al.\ 2005; Alexander et
al.\ 2005; Swinbank et al.\ 2008; Tacconi et al.\ 2008; Pope et al.\ 2008; 
Menendez-Delmestre et al.\ 2009).

The advent of the SPIRE instrument (Griffin et al.\ 2010) on board
\herschel\footnote{\herschel\ is an ESA space observatory with science
instruments provided by Principal Investigator consortia.  It is open
for proposals for observing time from the worldwide astronomical
community.} (Pilbratt et al.\ 2010) promises to shed new light on many
issues relating to submm galaxies. In photometry mode, SPIRE operates in three submm bands centred on 250, 350 and 500$\,\mu$m, and thus the large scale surveys conducted with SPIRE will detect many 10,000s of submm galaxies (e.g. Eales et al.\ 2010; Oliver et al., in prep). SPIRE observations of existing submm/mm survey fields provides the crucial SED information shortward of 500$\,\mu$m which cannot be routinely accessed from the ground (c.f. Kovacs et al.\ 2006), allowing us to characterise their IR luminosity (\lir) and dust temperature (\td) with unprecedented
accuracy (e.g.\ Chapman et al.\ 2010; Magnelli et al.\ 2010).

Armed with this new information we can re-evaluate the potential of using submm (and mm) wavelength data {\it alone} to estimate the redshifts of submm galaxies. As submm galaxies tend to be both high redshift ($<z>\sim2$; Chapman et al.\ 2005) and highly dust obscured they are extremely faint at optical/near-IR wavelengths and hence obtaining redshifts via optical spectroscopy is challenging. Because of this the prospects for submm/mm photometric redshifts have been discussed many times since the early SCUBA surveys (Blain 1999; Hughes et al.\ 2002; Aretxaga et al.\ 2003;  Blain, Barnard \& Chapman 2003; Pope \& Chary 2010; Schulz et al.\ 2010), with mixed results. The main obstacle for submm/mm photometric redshifts is that redshift and dust temperature are degenerate for the simplest, single temperature, modified blackbody SED. To break this degeneracy some other constraint is needed, such as the IR luminosity to dust temperature relation (Blain 1999; Blain, Barnard \& Chapman 2003) or information at other wavelengths, physically related to the far IR emission (e.g. radio or mid-IR; Carilli \& Yun 1999; Aretxaga et al.\ 2005; Aretxaga et al.\ 2007). 

In this paper we investigate the potential of submm/mm photometric redshifts using a sample selected at SPIRE and mm-wavelengths. We focus on two key survey fields; Great Observatories Origins Deep
Survey North (GOODS-N) and Lockman Hole North, which contain some of the deepest imaging on the sky at mm, submm, and radio wavelengths. In particular the Owen et al.\ (2008) and Morrison et al.\ (2010) VLA imaging of the Lockman Hole, and GOODS-N fields, represent the deepest `blank-field' images currently available at 1.4\,GHz. The depth of the radio imaging is crucial as without the positional information from identifications in high angular resolution radio imaging it is often not possible to correctly deblend the highly confused SPIRE imaging.

In \S\ref{sec:data}, we introduce the parent mm-wavelength catalogues utilised
in this work. \S\ref{sec:method} describes the process undertaken to find multiwavelength identifications for the mm and SPIRE sources, \S\ref{sec:spiredet} presents the SPIRE detection
statistics for our mm-wavelength sample, and describes the
properties of those sources with both mm and SPIRE detections. Finally,
\S\ref{sec:discussion} considers the scientific implications of these
results, and we present our conclusions in \S\ref{sec:conclusion}.
Throughout we assume a $\Lambda$CDM cosmology with
$\Omega_{\Lambda}=0.7$, $\Omega_{\rm m}=0.3$ and
$H_0=70$\,km\,s$^{-1}$\,Mpc$^{-1}$.

\section{Data}\label{sec:data}

The starting point for this study is the available mm-wavelength imaging in the GOODS-N and Lockman Hole North field.

In GOODS-N we make use of the combined MAMBO-AzTEC image and source catalogue of Penner et al.\ (2011). Combining the pre-existing 1.1-mm AzTEC (Perera et al.\ 2008) and MAMBO 1.2-mm (Greve et al.\ 2008) imaging of GOODS-N the resulting image has a typical noise of 0.5\,mJy at an effective wavelength of 1.16\,mm over an area of 0.08 deg$^2$. This combined image yields a list of 41 1.16-mm sources detected at a significance of $>3.8\sigma$ and  S$_{1.16mm}=2-10$\,mJy.
In Lockman Hole North we make use of the recent MAMBO 1.2-mm imaging and source catalogue of Lindner et al.\ (2011). Lindner et al.\ present a list of 41 1.2-mm sources detected with S/N$>4$ and S$_{1.2mm}=2-5$\,mJy.

In addition we make use of the SPIRE imaging at
250, 350 and 500\,$\mu$m obtained as part
of the Science Demonstration Phase (SDP) of \herschel\ by the
\herschel\ Multi-Tiered Extragalactic Survey (HerMES\footnote{hermes.sussex.ac.uk} -- Oliver et al.,
in prep.).  The SPIRE instrument, its in-orbit performance and
its scientific capabilities are described by Griffin et al.\ (2010);
its calibration methods and accuracy are outlined in Swinyard et al.\
(2010). We use images produced as described in
Levenson et al.\ (2010). The SPIRE instrument has a FWHM of 18, 25 and 36 arcsec at 250, 350 and 500$\,\mu$m, respectively. HerMES imaging of GOODS-N and Lockman Hole North has a typical 5-$\sigma$ depth at all SPIRE wavelengths of $\sim$5\,mJy and $\sim$10\,mJy, respectively, ignoring the dominant contribution from
confusion (Smith et al., in press).

At radio wavelengths we make use of the VLA 1.4-GHz source catalogue from Morrison
et al.\ (2010) and Owen et al.\ (2008) for the GOODS-N and Lockman Hole fields, respectively. In GOODS-N, the Morrison et al.\ (2010) imaging have a sensitivity of
3.9$\,\mu$Jy\,beam$^{-1}$ with the sensitivity declining to 8$\,\mu$Jy\,beam$^{-1}$ at a
radius of 15 arcmin from the centre. In Lockman Hole North, the Owen et al.\ (2008) imaging has a sensitivity of 2.7$\,\mu$Jy\,beam$^{-1}$ over the central 40 arcmin $\times$ 40 arcmin region overlapping the mm and SPIRE observations.

%, partly due to the $\sim$30-arcmin {\sc fwhm} primary beam,
%partly because of bandwidth smearing.
%At 850$\,\mu$m we make use of the so-called `SCUBA supermap' of
%GOODS-N (Borys et al.\ 2003, Pope et al.\ 2006, Wall, Pope \& Scott 2008), using the flux densities extracted from
%a combined analysis of the 850-$\mu$m and 1.2-mm maps by G08.

%In the optical to mid-IR regime, we make use of the multi-wavelength
%observations taken as part of the GOODS programme, including the
%\spitzer\ IRAC and MIPS 24-$\mu$m catalogues of Magnelli et al.\
%(2009), and \hubble\ ACS catalogue of Giavalisco et al.\ (2004). These data are among the deepest available, with typical
%5-$\sigma$ noise levels of $\sim$1 and $\sim$5\,$\mu$Jy at 3.6 and
%24$\,\mu$m, respectively; the optical data have a typical 5-$\sigma$
%point-source detection limit of $m_{\rm B}<27$ (AB mag).

\section{Multiwavelength Identifications for submm/mm detected sources}\label{sec:method}
The large beam size ($>10$ arcsec {\sc FWHM}) of typical submm/mm imaging facilities means that accurately compiling multiwavelength identifications and photometry for submm/mm detected sources is often a significant challenge (e.g. Lilly et al.\ 1999; Ivison et al.\ 2007; Roseboom et al.\ 2009; Chapin et al.\ 2011). This challenge arises mainly in two ways; firstly the large beam size, and low signal-to-noise, of typical submm/mm surveys means that catalogued sources will have quite large positional uncertainties. This makes matching to catalogues at other wavelengths difficult, as there may be more than one potential counterpart within the positional errors of the submm/mm source. Indeed, for faint submm sources at high redshift, the source density of matching catalogues deep enough to contain the true match may be so high that more than one match can always be found.

The second problem is source confusion; more than one astronomical source is present in the submm/mm beam, and hence catalogued sources are made up of multiple astronomical objects. This problem is particularly worrying as it affects the integrity of the submm/mm source catalogues themselves.

Several statistical methods which can account for one or both of these issues exist (Downes et al.\ 1986; Chapin et al.\ 2011; Roseboom et al.\ 2009; Roseboom et al.\ 2010). For the datasets considered here one or both of these issues are present. Source identification is a major issue for the mm-wavelength datasets, but source confusion is not, as mm-sources have large positional uncertainties, but low source densities. The SPIRE datasets have both large positional uncertainties and high source densities, and so are equally affected by both. 

Thus different approaches are needed to reliably produce multi-wavelength associations for the different submm/mm datasets. In practice this means that it is generally not possible to reliably identify SPIRE counterparts to mm-detected sources without first making identifications in some other catalogue with higher spatial resolution (i.e. 1.4\,GHz or 24$\,\mu$m). To emphasise this point; the typical flux density of our mm-detected sample, after correcting for flux boosting,  is $S_{1.2mm}\sim2$ mJy. Taking rough estimates of the mean dust temperature and redshift of submm/mm sources; 35\,K (Kovacs et al.\ 2006; Chapman et al.\ 2010) and $z=2.5$ (Wardlow et al.\ 2011), and assuming a single temperature modified blackbody SED, gives predicted SPIRE fluxes of 13, 19 and 15 mJy at 250, 350 and 500\,$\mu$m, respectively. These flux levels are below the confusion `limit' at these wavelengths, with the best estimate of the noise from confused sources $\sigma_{\rm conf}=6$\,mJy  (Nguyen et al. 2010; Glenn et al. 2010). Hence recovering the SPIRE counterparts to mm-detected sources will require a technique which can reduce the effect of confusion, by using prior information at other wavelengths (Roseboom et al.\ 2010, Chapin et al.\ 2011). Putting the issue of source confusion aside, even if these sources could be recovered reliably, source identification would still pose a serious challenge. The positional uncertainties of both the mm and SPIRE detected sources ($\sigma_{\rm pos}\sim 2$--3 arcsec) requires the use of a 9--12 arcsec matching radius in order to recover the bulk of the true mm-SPIRE associations. Given known number density of SPIRE sources at these flux levels ($\sim$ 7000 deg$^{-2}$; Glenn et al.\ 2010) we would expect between 14--25 per cent of these associations to be chance alignments.

In this work we first match the mm-detected source lists to 1.4\,GHz sources, using the well established catalogue based cross-matching techniques of Downes et al. (1986) and then use these 1.4\,GHz source positions, along with the source positions of all known 1.4\,GHz and 24$\,\mu$m sources in these fields, to \lq deblend\rq\ the SPIRE maps into their individual contributors and thus provide the best estimate of the SPIRE photometry at the locations of our mm-selected sample.

\subsection{1.4 GHz identifications of mm-selected sources}\label{sec:radioids}

In the GOODS-N field we search for potential 1.4\,GHz counterparts in the Morrison et al.\ (2010) VLA source catalogue to the 41 1.16-mm sources from Penner et al.\ (2011) using a 10 arcsec search radius. 28 of the 41 sources have at least one potential counterpart within this search radius. To determine the reliability of these matches, we estimate the probability of a
chance alignment using the $P$-statistic, as defined by Downes et al.\
(1986).  Defining a reliable match to be one with a less than 5 per cent probability of being a chance alignment, 32 reliable matches are found for 24 objects, with 8 sources having more than one reliable radio counterpart.

For the sources with more than one reliable counterpart we take the counterpart with the lower $P$-statistic to be the correct identification. 

%As with any matching there are a number of notable cases which need special attention. GOODS-N is one of the most well-studied regions of the sky at all wavelengths including the submm/mm, and our work here is not the first attempt to make associations between mm-detected sources and radio observations, so care is needed in the case of a few well-known objects.

One difficult case for our matching is HDF 850.1 (ID 14 in the Penner et al.\ 2011 catalogue), the brightest submm source detected in the original 850$\,\mu$m SCUBA imaging of GOODS-N (Hughes et al.\ 1998). Our naive cross matching approach would associate this object with a 56$\,\mu$Jy radio source located 5.3 arcsec away. However subsequent near-IR and high resolution Submillimetre Array (SMA) observations show that the true position is much closer to the original SCUBA position (Dunlop et al.\ 2004; Cowie et al.\ 2009), although no distinct 1.4\,GHz radio source can be found near this position in the Morrison et al.\ (2010) image. Here we adopt the Dunlop et al.\ (2004) position for matching with the SPIRE data.

Other difficult cases are those sources for which the submm emission is later shown, via high resolution mm-wavelength interferometry, to originate from multiple components. In these cases the 1.4\,GHz identification(s) is often found to lie co-incident with one (or more) of these components (e.g. Wang et al.\ 2011). Amongst our sample are two examples of this, GN20 (G1 in our sample; Daddi et al.\ 2009a; Carilli et al.\ 2010; Carilli et al.\ 2011) and GN21 (G19 in our sample; Wang et al.\ 2011). In both cases we have simply taken a single radio identification, the one with the lowest $P$-statistic, as being solely responsible for the mm-flux density. For G1 the radio identification corresponds to GN20 from Daddi et al.\ 2009a, while for G19 this corresponds to GOODS 850-13c in the list of Wang et al.\ 2011. While it may turn out that a large fraction of mm-detected sources do indeed contain multiple components, in both G1 and G19 the single radio identification we have chosen is found to be responsible for $\gs 50$ of the submm/mm flux (Daddi et al.\ 2009a; Wang et al.\ 2011). Importantly, our SPIRE photometry pipeline accounts for all 1.4\,GHz, mm, and 24$\,\mu$m detected sources, so it should not be affected by this issue.

Encouragingly, for the 28 sources from Penner et al.\ (2010) which were previously identified in the AzTEC and/or MAMBO surveys we recover the same identifications made by both Chapin et al.\ (2009), for the 25 sources in common, and Greve et al.\ (2008), for the 8 sources in common.

%Interestingly this identification is the same as the one made in Chapin et al.\ 2009 for the AzTEC counterpart to GN1200.9; AzGN28. As we will see in Section~\ref{sec:detwithid} this identification is also supported by the SPIRE--MAMBO photometry.

In the Lockmann Hole North field we make use of the existing matches between the 1.2-mm sources and 1.4 GHz radio catalogue presented in Lindner et al.\ (2011). 40 of the 41 1.2-mm sources are found to have a 1.4 GHz counterpart within 8 arcsec, with 39 of these found to have less than a 5 per cent probability of being a chance alignment and hence are deemed reliable.

It is worth noting that 3 of these reliable radio counterparts are not found in the published Owen et al.\ (2010) 1.4\,GHz source lists, and come from a deeper extraction performed in the vicinity of the MAMBO sources by Lindner et al.\ (2011). For these sources we adopt the quoted 1.4\,GHz positions and flux densities from Lindner et al.\ (2011).

Thus we are left with a sample of 63 mm-selected sources with high angular resolution 1.4 GHz positions with which to match to the SPIRE data.% While we do not make use of the remaining 18 sources without reliable radio counterparts we can speculate as to the nature of these sources, and in particular

Interestingly there is a significant disparity between the identification rates in the 2 fields considered here. Both Penner et al.\ (2010) and Lindner et al.\ (2011) list 41 mm-detected sources with a similar noise; $1\sigma\sim0.7$\,mJy, although in both cases the noise varies significantly across the field. However the effective area of the two surveys is quite different; The Penner et al.\ (2010) combined AzTEC/MAMBO image of GOODS-N covers an area of 0.08 deg.$^2$, while the Lindner et al.\ (2011) MAMBO image of Lockman North covers an area of 0.16 deg.$^2$. Thus it is clear that for the Penner et al.\ GOODS-N sample to find the same number of objects in similar to one half the area it must be identifying a fainter population of sources. The differential number counts at these wavelengths are known to be well described by a power law $dN/dS\propto S^{\delta}$, with an exponent $\delta\sim-3$. So to observe a difference of a factor of 2 in surface density, the GOODS-N sample must be effectively a factor of $\sqrt2$ deeper, ignoring the $k$-correction between 1.16 and 1.2 mm.

\subsection{SPIRE Photometric method}

The most difficult obstacle to determining accurate SPIRE photometry is the effect of source confusion, i.e.\ the
contributions of numerous faint sources within a single SPIRE
resolution element, centred on the target of interest. For this reason, several authors have developed techniques
that utilise the positions of sources detected at other wavelengths,
usually 24\,$\mu$m and 1.4\,GHz, to disentangle the various
contributions from discrete sources to the SPIRE flux in a given beam
element (e.g. Chapin et al.\ 2010, Roseboom et al.\ 2010). This process is
made possible by the high correspondence between the 24-$\mu$m and
1.4-GHz populations and those observed at Far-IR wavelengths; $>80$ per cent of the cosmic IR background (CIB) at SPIRE
wavelengths can be accounted for by 24-$\mu$m sources with
$S_{24}>25\,\mu$Jy (Pascale et al.\ 2009; Marsden et al.\ 2009), while the strong 
correlation between the Far-IR and radio luminosity is known to hold across a wide range in redshift and luminosity (e.g.\ Ibar et al.\ 2008; Ivison et al.\ 2010a).

 Here we follow the
prescription presented by Roseboom et al.\ (2010; hereafter R10) to
measure the SPIRE fluxes of mm sources in GOODS-N and Lockman Hole North, with a number of
modifications. Below, we summarise the R10 methodology, then highlight
our departures from it.

R10 assume that the SPIRE map can be fit directly by assuming that the positions of all sources contributing significantly to the map are known (i.e. previously detected at 24\,$\mu$m, 1.4\,GHz, or other wavelengths), and that only the SPIRE flux density of each of these sources is unknown. Algebraically this means the observed SPIRE map, $d$, can be described as a series of
point sources with fluxes, $f$, at known positions, $x$. Thus, the
observed map can be described as \[d=P_xf+\delta,\] Where $P_x$ is the
point response function (PRF) centred at position $x$ and $\delta$
is some unknown noise term. As discussed in R10, this equation can be
easily inverted, using linear algebra, to obtain the best-fit values
for the fluxes. However, one potential pitfall of this approach is that of
over-fitting. If many of our input sources are intrinsically faint and
the source density is high, this approach will give poor results. R10
solve this issue by using model-selection techniques to eliminate
those sources that are unnecessary to describe the map.  Specifically
the Akiake Information Criterion (AIC; Akiake 1974) is used to filter the input list of
sources via backward elimination. This is very time-consuming and
hence the number of sources that could be fitted simultaneously was
limited to $\sim$100--200 in R10.

Here, we replace this model-selection approach with the adaptive Least
Absolute Shrinkage and Selection Operator (LASSO -- Tibshirani 1996;
Zou 2006). In particular, we combine a non-negative application of
this technique -- the Non-Negative Least Squares path (NNLSpath)
algorithm of ter Braak et al.\ (2010) -- with the adaptive weighting
scheme suggested by Zou (2006). In Zou (2006), the weights are
tunable, based on initial estimates of the parameters; here we
use the probability of a chance alignment, $\phi$, as defined in
R10. Specifically, $\phi$ is calculated separately for the three samples which make up our input list; radio, 24$\,\mu$m and mm-detected sources. In cases where a source is detected in more than one of these bands the lowest value of $\phi$ is adopted. As mm-detected sources have the lowest areal density these typically have higher weightings, followed by radio sources, and then 24$\,\mu$m. 

Accuracy in model selection is the primary advantage, the
adaptive LASSO is known to have `oracle' properties\footnote{In
model selection, a method is said to have oracle properties if it can
reliably identify which free parameters are truly present in the underlying
system, and which are unnecessary or spurious.} with the correct choice of
weightings, however there is also a significant advantage in computational
speed. With this approach we can solve for large numbers of candidate
sources (up to 10,000) in a realistic amount of time.

In addition, we modify the background estimation. Rather than use an
unconstrained local estimate of the background, the background is estimated globally, assuming a flat background across the total extent of the map. This approach is prefered as it is more robust against incompleteness in the prior input list, and because there are no significant large scale correlations in the noise present in our SPIRE maps due to the exceptionally low frequency ``knee'' to the $1/f$ noise (Griffin et al.\ 2010).

Finally potential new SPIRE sources, i.e. those that were previously unknown at 24$\,\mu$m or radio wavelengths, are considered by examining the residual maps after fitting the SPIRE fluxes of known sources.

\subsection{Estimating the true noise for SPIRE photometry method}
Accurately estimating the effect of confusion on an individual source is non-trivial. As the SPIRE maps have a mean intensity of zero, for blind source catalogues we can expect that the typical confusion noise is given by the width of the distribution of flux densities (P(D)) in a PRF-convolved image. For SPIRE we know these values are 5.8, 6.3 and 6.8 mJy/beam at 250, 350 and 500$\,\mu$m, respectively (Nguyen et al.\ 2010). However our source extraction process both attempts to find the true zero point of the map, and simultaneously fits the positions of sources found at other wavelengths, in an attempt to reduce this noise. But it is difficult to assess for a given source how much we have `resolved' the confusing background, and how much is left as unresolved fluctuations in the map.

One simple way to assess the global improvement found by our method is to use completely synthetic maps and catalogues produced using input mock catalogues known to give reasonable agreement to the known far-IR and submm number counts (e.g. Fernandez-Conde et al.\ 2008). Testing against simulated SPIRE maps with the same properties as our GOODS-N data (i.e. instrumental noise, number counts and pixel intensity distribution) we find a typical {\it total} noise, i.e. confusion and instrumental noise, of $\sigma_{\rm total}$ = 2.8, 3.9, 3.7\,mJy at 250, 350 and 500$\,\mu$m, respectively. These values compare quite favourably to the SPIRE photometry process described in R10, which gave $\sigma_{\rm total}$ = 4.9, 5.4, 6.5\,mJy at 250, 350 and 500$\,\mu$m for the same simulations.

While testing the performance of our method on simulated datasets is of interest, we really need an accurate estimate of the true error on the SPIRE photometry of our actual sources. One approach would be to estimate the global confusion noise in our SPIRE photometry by investigating residual intensity fluctuations in the SPIRE maps, {\it after} subtracting the contributions from `resolved' sources. This approach is appealing, as we know that the residual maps contain information about sources not included in our prior source list, as well as any artefacts from incorrect astrometry and/or assumptions about the PRF.

However there is one complication for this approach; variations in sensitivity in the prior source lists across the SPIRE map will mean the ability to resolve confusing sources varies spatially. Thus we must assess the confusion noise on scales small enough that the prior source list sensitivity is invariant. This is particularly difficult in this work, as the combination of 24$\,\mu$m and 1.4\,GHz prior source catalogues results in large variations across the field. This is due to both the difference in relative sensitivity of 24$\,\mu$m and 1.4\,GHz to SPIRE sources, as well as the tapering of the 1.4\,GHz sensitivity towards the edges of its coverage.

We utilise a variant of this approach to estimate the true error for our confused SPIRE sources here. Specifically we construct PRF convolved residual maps for each of the SPIRE bands, using our best estimate of the SPIRE photometry at the point of each input source and the known SPIRE beam. Then at the position of each source we are interested in we assess the total noise per pixel by measuring the standard deviation of surrounding pixels within a 20 pixel radius (2, 2.8 and 4 arcmin for the SPIRE 250, 350 and 500$\,\mu$m maps used here). 

A further complication arises in converting this noise per pixel into the noise for a point source as we know that the confusing background must be correlated, on small scales by the PRF, and on larger scales by the intrinsic clustering of the sources. This means that in calculating the total noise estimate for a given point source we must take into account the uncorrelated instrumental noise in each pixel, and the correlated noise from confusion. For simplicity we assume that the correlated noise on small scales is dominated by the PRF, and assume the shape of the PRF gives the covariance between pixels directly, i.e. If we assume the SPIRE photometry is estimated by 
\[S=\frac{\sum_i d_iP_i^2\sigma_i^2}{\sum_i P_i^2/\sigma_i^2},\]

where $d_i$ is the intensity in pixel $i$, $P_i$ the PRF for the point source in pixel $i$ and $\sigma_i$ the instrumental noise in pixel $i$, the noise estimate, including contributions from both confusion noise and instrumental noise can be given by
\[\sigma_{\rm total}^2=1/\left({\bf P}^T{\bf C}^{-1}{\bf P}\right),\]
where {\bf C} is the covariance matrix, the diagonal elements of which are given by 1/$\sigma_i^2$, and the off diagonal elements $i,j$ given by $P'_j/\sigma_i/\sigma_j$, where {\bf P'} is the PRF assuming a point source centred at position $i$.

Using this approach we find the typical 1$\sigma$ total noise for our mm-detected sources is $\sigma_{\rm total}=$2.2, 2.6 and 2.5 mJy in the central part of the GOODS-N field for the 250, 350 and 500$\,\mu$m bands, respectively, increasing to 3.7, 4.8 and 5.7 mJy at a radius of 20 arcmin. Similarly in the Lockman North field the total noise is found to be 3.5, 3.6 and 4.7 mJy at the centre of the field, and 4.4, 4.7, 6 mJy at a radius of 20 arcmin.

It is worth noting that these variations are solely due to the variations in the depth of the prior source list, the SPIRE sensitivity is uniform across both fields within the coverage of the mm observations (Levenson et al.\ 2010).

\section{Results}
\label{sec:spiredet}
\subsection{SPIRE detection statistics}

We define the SPIRE measurement as a detection if it has $>$3-$\sigma$ significance in the flux
density estimate and the goodness of the photometric fit to the map is high 
(reduced $\chi^2<5$, in a $15\times15$ pixel region centred on the source position). The $\chi^2$ cut is necessary to remove cases where
bright or unaccounted for sources, or residual non-Gaussian noise (e.g.\
Galactic cirrus, or uncorrected thermal drifts) introduce signal that
cannot be accounted for by our input source model.

Table~\ref{tab:gnspirephot}
presents our SPIRE photometry for all mm-selected sample in GOODS-N, while Table~\ref{tab:lhnspirephot} presents SPIRE photometry for mm-sources in the Lockman Hole North field.%; postage-stamp images for the 24--500-$\mu$m bands are shown in Appendix~\ref{sec:stamps}.

\begin{table*}
\caption{Measured SPIRE photometry for mm-sources in GOODS-N. 3$\sigma$ upper limits are given for sources which are not detected above this significance in the SPIRE imaging. For the mm photometry deboosted flux densities are quoted. The $\chi^2$ represents the reduced $\chi^2$ in a $15\times15$ pixel region centred on the source position. ID number refers to the equivalent value from Penner et al.\ (2010). C09 refers to the ID numbers of sources in common with Chapin et al.\ (2009). G08 refers to the ID numbers of sources in common with Greve et al.\ (2008). 1.4\,GHz flux density is from the catalogue of Morrison et al.\ (2010).}%Redshifts, either spectroscopic or robust optical/near-IR photometric estimates, from the literature are quoted where available.}
\begin{scriptsize}
\label{tab:gnspirephot}
\begin{tabular}{lllllllllllll}

%\hline
%\multicolumn{8}{l}{GOODS North}\\
\hline
Name & C09$^{\dagger}$ & G08$^{\ddagger}$ & R.A. & Dec. & $S_{250}$ & $\chi^2_{250}$ & $S_{350}$ & $\chi^2_{350}$ & $S_{500}$ & $\chi^2_{500}$ & $S_{\rm 1.2mm}^{\star}$  &  $S_{\rm 1.4 GHz}$ \\%& Redshift\\
 & & & (deg) & (deg) & (mJy) & & (mJy) & &  (mJy) & & (mJy) & (mJy)\\
\hline\hline
G1$^*$ & 1 & 1 & 189.29950 & 62.37013 & 18.6 $\pm$ 2.3 & 0.97 & 31.7 $\pm$ 2.7 & 2.16 & 24.3 $\pm$ 2.7 & 1.10 & 10.5$\pm$0.7 & 0.079 $\pm$ 0.014 \\%&       4.05$^a$ \\
G2$^*$ & 3 & 2 &189.13918 & 62.23579 & 18.1 $\pm$ 2.3 & 1.77 & 15.2 $\pm$ 2.7 & 2.11 & $<$ 7.8 & 1.68 & 5.2$\pm$0.6 & 0.036 $\pm$ 0.004 \\%&       4.05$^b$ \\
G3$^*$ & 5 & 4 &189.37811 & 62.21633 & 46.3 $\pm$ 2.0 & 1.25 & 46.5 $\pm$ 2.3 & 1.75 & 32.7 $\pm$ 2.3 & 1.27 & 4.4$\pm$0.6 & 0.123 $\pm$ 0.004 \\%&       2.98$^c$ \\
G4$^*$ & 7 & 3 &189.29727 & 62.22540 & 38.6 $\pm$ 2.1 & 1.76 & 33.0 $\pm$ 2.4 & 2.62 & 23.2 $\pm$ 2.3 & 1.30 & 4.0$\pm$0.5 & 0.131 $\pm$ 0.009 \\%&       1.996$^d$ \\
G5 & 2 & 13  &189.13301 & 62.28749 & $<$ 6.9 & 1.58 & $<$ 8.7 & 1.68 & $<$ 7.8 & 0.88 & 4.1$\pm$0.6 & 0.026 $\pm$ 0.005 \\%&            \\
G6$^*$ & 6  &5 &189.11346 & 62.10158 & 9.2 $\pm$ 2.4 & 1.23 & 13.5 $\pm$ 3.0 & 1.61 & 5.2 $\pm$ 2.9 & 1.14 & 4.5$\pm$0.7 & 0.053 $\pm$ 0.012 \\%&            \\
G7 & 4 &12 &188.95979 & 62.17839 & $<$ 7.5 & 1.53 & $<$ 8.1 & 2.31 & $<$ 8.4 & 2.09 & 4.6$\pm$0.7 & 0.028 $\pm$ 0.005 \\%&            \\
G8$^*$ & 26 &6 &189.30772 & 62.30739 & 8.2 $\pm$ 2.2 & 1.91 & 8.1 $\pm$ 2.8 & 3.76 & $<$ 7.8 & 2.94 & 3.1$\pm$0.6 & 0.652 $\pm$ 0.004 \\%&            \\
G9$^*$ & 11 &14 &189.14937 & 62.11888 & 33.9 $\pm$ 2.1 & 1.45 & 20.3 $\pm$ 2.7 & 1.16 & $<$ 8.4 & 1.43 & 2.9$\pm$0.6 & 0.046 $\pm$ 0.008 \\%&      0.952$^e$ \\
G11$^*$ & 13& 15& 188.97157 & 62.22708 & 21.0 $\pm$ 2.4 & 1.25 & $<$ 9.3 & 4.06 & $<$ 8.4 & 2.39 & 2.6$\pm$0.6 & 0.045 $\pm$ 0.005 \\%&       2.098$^c$ \\
G13 & 12 & &189.13604 & 62.10586 & $<$ 6.9 & 1.44 & $<$ 8.7 & 1.35 & $<$ 8.7 & 1.25 & 2.7$\pm$0.7 & 0.027 $\pm$ 0.005 \\%&            \\
G14 & 14 &25 &189.21696 & 62.20714 & $<$ 6.3 & 1.05 & $<$ 7.5 & 1.77 & $<$ 7.2 & 1.31 & 2.4$\pm$0.6 & 0.016 $\pm$ 0.004$^1$ \\%&            \\
G16 &  & &189.20117 & 62.35201 & $<$ 9 & 1.00 & $<$ 9.3 & 1.98 & $<$ 8.1 & 1.24 & 2.5$\pm$0.6 & 0.030 $\pm$ 0.005 \\%&            \\
G18$^*$ &16  & &189.06708 & 62.25388 & $<$ 7.2 & 1.35 & $<$ 9.3 & 2.95 & 14.1 $\pm$ 2.6 & 1.46 & 2.1$\pm$0.6 & 0.036 $\pm$ 0.005 \\%&       2.578$^d$ \\
G19$^*$ &  &29 &189.30020 & 62.20341 & 11.2 $\pm$ 2.1 & 1.50 & 11.3 $\pm$ 2.4 & 1.28 & 17.3 $\pm$ 2.4 & 1.09 & 2.1$\pm$0.6 & 0.032 $\pm$ 0.004 \\%&       2.914$^d$ \\
G21$^*$ & 9 & &189.40941 & 62.29355 & 7.1 $\pm$ 2.0 & 1.10 & 12.1 $\pm$ 2.6 & 1.53 & $<$ 8.4 & 1.01 & 2.1$\pm$0.6 & 0.026 $\pm$ 0.006 \\%&       3.19$^d$ \\
G24 & 24 & &189.03583 & 62.24325 & $<$ 7.5 & 1.34 & $<$ 9.6 & 3.08 & $<$ 8.1 & 2.36 & 2.0$\pm$0.6 & 0.039 $\pm$ 0.004 \\%&            \\
G26$^*$ & 15 & &188.94971 & 62.25811 & $<$ 8.1 & 2.33 & $<$ 10.2 & 4.22 & 10.1 $\pm$ 2.9 & 1.93 & 2.1$\pm$0.7 & 0.124 $\pm$ 0.006 \\%&            \\
G27$^*$ & 18 & &189.42151 & 62.20591 & 31.0 $\pm$ 2.3 & 1.19 & 27.9 $\pm$ 2.4 & 0.99 & 30.3 $\pm$ 2.5 & 0.88 & 1.9$\pm$0.6 & 0.033 $\pm$ 0.005 \\%&            \\
G29$^*$ & 25  & &189.21550 & 62.08417 & 12.6 $\pm$ 2.9 & 1.78 & 15.1 $\pm$ 3.3 & 2.57 & 17.7 $\pm$ 3.0 & 1.43 & 2.6$\pm$0.9 & 0.079 $\pm$ 0.006 \\%&            \\
G30$^*$ & 17 & &188.92167 & 62.24403 & 11.7 $\pm$ 2.6 & 1.59 & $<$ 10.2 & 3.80 & $<$ 9. & 2.27 & 2.2$\pm$0.7 & 0.029 $\pm$ 0.005 \\%&            \\
G31$^*$ &  & &189.32254 & 62.13453 & 35.1 $\pm$ 2.5 & 3.20 & 32.1 $\pm$ 3.0 & 1.95 & 22.5 $\pm$ 2.7 & 0.96 & 2.0$\pm$0.7 & 0.050 $\pm$ 0.008 \\%&            \\
G34 &  & &189.14537 & 62.32322 & $<$ 7.2 & 1.36 & $<$ 8.7 & 2.28 & $<$ 10.3 & 0.99 & 1.8$\pm$0.6 & 0.085 $\pm$ 0.005 \\%&            \\
G41$^*$ &  & &189.55217 & 62.24859 & 25.3 $\pm$ 3.2 & 2.48 & 25.6 $\pm$ 3.5 & 9.72 & 15.3 $\pm$ 3.7 & 2.97 & 1.9$\pm$0.7 & 0.056 $\pm$ 0.006 \\%&            \\

\hline
\multicolumn{9}{l}{$^{\dagger}$ Typically written with prefix AzGN}\\
\multicolumn{9}{l}{$^{\ddagger}$ Typically written with prefix GN1200.}\\
\multicolumn{9}{l}{$^{\star}$Deboosted flux density as described in Penner et al.\ (2011)}\\
\multicolumn{9}{l}{$^*$ detected in at least 1 SPIRE band}\\
\multicolumn{9}{l}{$^{1}$ 1.4\,GHz flux density from Dunlop et al.\ 2004}\\

\end{tabular}
\end{scriptsize}
\end{table*}
\begin{table*}
\caption{Measured SPIRE photometry for mm-sources in Lockman Hole North. 3$\sigma$ upper limits are given for sources which are not detected above this significance in the SPIRE imaging. For the mm photometry deboosted flux densities are quoted. The $\chi^2$ represents the reduced $\chi^2$ in a $15\times15$ pixel region centred on the source position. ID number refers to the equivalent value from Lindner et al.\ (2011). 1.4\,GHz flux density is from the catalogue of Owen  et al.\ (2008), as presented by Lindner et al.\ (2011).}%Redshifts, either spectroscopic or robust optical/near-IR photometric estimates, from the literature are quoted where available.}
\begin{scriptsize}
\label{tab:lhnspirephot}
\begin{tabular}{lllllllllll}

%\hline
%\multicolumn{8}{l}{Lockman Hole North}\\
\hline
ID & R.A. & Dec. & $S_{250}$ & $\chi^2_{250}$ & $S_{350}$ & $\chi^2_{350}$ & $S_{500}$ & $\chi^2_{500}$ & $S_{\rm 1.2mm}$ (deboosted)  & $S_{\rm 1.4 GHz}$ \\%& Redshift\\
 & (deg) & (deg) & (mJy) & & (mJy) & &  (mJy) & & (mJy) & (mJy)\\
\hline\hline
   L1$^*$ & 161.75083 & 59.01878 & 80.1 $\pm$ 3.4 & 0.92 & 63.2 $\pm$ 3.8 & 0.85 & 40.6 $\pm$ 4.4 & 0.89 & 3.5$^{+0.6}_{-0.6}$ & 0.279 $\pm$ 0.015\\%&2.562$^a$\\
   L2$^*$ & 161.61192 & 59.09578 & $<$ 11.7 & 1.07 & 12.5 $\pm$ 4.0 & 0.99 & $<$ 13.8 & 1.08 & 3.8$^{+0.7}_{-0.7}$ & 0.036 $\pm$ 0.006\\%&\\%%4.29$\pm1.33^b$\\
   L3 & 161.63112 & 58.84889 & $<$ 10.8 & 0.84 & $<$ 11.1 & 0.92 & $<$ 13.2 & 0.87 & 3.8$^{+0.8}_{-0.7}$ & 0.459 $\pm$ 0.025\\%&\\
   L4 & 161.53050 & 58.90389 & $<$ 10.8 & 0.83 & $<$ 11.4 & 1.03 & $<$ 14.1 & 0.79 & 2.7$^{+0.5}_{-0.5}$ & 0.030 $\pm$ 0.004\\%&\\
   L5$^*$ & 161.85592 & 59.06014 & 32.8 $\pm$ 3.9 & 1.21 & 44.5 $\pm$ 4.1 & 1.08 & 35.6 $\pm$ 4.6 & 0.92 & 4.0$^{+0.8}_{-0.9}$ & 0.051 $\pm$ 0.006\\%&3.$\pm0.42^b$\\
   L6$^*$ & 161.66112 & 58.93681 & 52.4 $\pm$ 3.6 & 0.91 & 51.5 $\pm$ 3.6 & 0.75 & 17.5 $\pm$ 4.4 & 2.17 & 2.3$^{+0.4}_{-0.4}$ & 0.160 $\pm$ 0.010\\%&2.03$^{c}$\\
   L7$^*$ & 161.75054 & 58.91150 & 21.0 $\pm$ 3.6 & 0.84 & 18.6 $\pm$ 3.5 & 0.97 & 31.7 $\pm$ 4.4 & 0.78 & 2.3$^{+0.4}_{-0.5}$ & 0.042 $\pm$ 0.005\\%&\\
   L8$^*$ & 161.63779 & 58.86642 & 30.0 $\pm$ 3.6 & 0.99 & 29.8 $\pm$ 3.7 & 0.94 & 16.5 $\pm$ 4.4 & 1.25 & 2.7$^{+0.6}_{-0.6}$ & 0.098 $\pm$ 0.010\\%&\\
   L9 & 161.77042 & 58.83556 & $<$ 10.8 & 0.96 & $<$ 11.1 & 0.91 & $<$ 13.8 & 1.21 & 3.8$^{+1.0}_{-0.9}$ & 0.023 $\pm$ 0.005\\%&\\
   L10$^*$ & 161.59604 & 58.99300 & 13.3 $\pm$ 3.5 & 0.72 & $<$ 14.1 & 0.94 & 19.8 $\pm$ 4.4 & 0.91 & 2.4$^{+0.5}_{-0.5}$ & 0.078 $\pm$ 0.007\\%&\\
   L11$^*$ & 161.48721 & 58.88856 & 58.2 $\pm$ 3.7 & 1.02 & 56.6 $\pm$ 3.8 & 1.26 & 18.8 $\pm$ 4.6 & 1.44 & 2.7$^{+0.6}_{-0.6}$ & 0.315 $\pm$ 0.019\\%&1.95$^{c}$\\
   L12$^*$ & 161.19829 & 59.00997 & 25.3 $\pm$ 3.7 & 0.84 & 36.0 $\pm$ 3.9 & 1.11 & 24.4 $\pm$ 4.7 & 0.78 & 2.7$^{+0.6}_{-0.7}$ & 0.273 $\pm$ 0.019\\%&2.16$\pm0.24^b$\\
   L13$^*$ & 161.53646 & 58.97458 & 12.1 $\pm$ 3.6 & 0.98 & 17.4 $\pm$ 3.6 & 0.80 & $<$ 14.1 & 0.82 & 2.1$^{+0.5}_{-0.5}$ & 0.198 $\pm$ 0.010\\%&\\%1.14$\pm0.48^b$\\
   L14$^*$ & 161.64937 & 59.13014 & 39.0 $\pm$ 3.9 & 0.99 & 46.2 $\pm$ 4.0 & 1.07 & 33.9 $\pm$ 4.6 & 1.65 & 3.0$^{+0.9}_{-0.9}$ & 0.097 $\pm$ 0.006\\%&2.26$\pm0.45^b$\\
   L15$^*$ & 161.86654 & 58.87058 & 55.8 $\pm$ 3.9 & 1.01 & 69.1 $\pm$ 3.6 & 0.95 & 49.7 $\pm$ 4.8 & 0.80 & 3.0$^{+0.9}_{-1.0}$ & 0.180 $\pm$ 0.018\\%&2.76$\pm0.18^b$\\
   L16$^*$ & 161.83633 & 58.86472 & $<$ 11.7 & 1.04 & 11.2 $\pm$ 3.6 & 0.90 & $<$ 14.1 & 0.95 & 2.7$^{+0.8}_{-0.8}$ & 0.048 $\pm$ 0.007\\%&\\
   L17$^*$ & 161.54250 & 59.04508 & 10.4 $\pm$ 3.7 & 1.15 & 14.6 $\pm$ 3.9 & 1.26 & $<$ 13.5 & 0.84 & 2.0$^{+0.5}_{-0.5}$ & 0.027 $\pm$ 0.003\\%&\\
   L18$^*$ & 161.73021 & 58.83444 & 39.9 $\pm$ 3.6 & 0.81 & 37.8 $\pm$ 3.7 & 0.91 & 25.4 $\pm$ 4.6 & 0.95 & 2.9$^{+1.1}_{-1.1}$ & 0.105 $\pm$ 0.008\\%&\\%1.3$\pm0.3^b$\\
   L19 & 161.25833 & 59.06758 & $<$11.4 & 0.91 & $<$ 12.3 & 1.23 & $<$ 14.7 & 1.45 & 2.0$^{+0.6}_{-0.6}$ & 0.029 $\pm$ 0.004\\%&\\
   L21$^*$ & 161.37575 & 59.11008 & $<$ 11.7 & 1.02 & 12.0 $\pm$ 4. & 1.11 & 16.0 $\pm$ 4.9 & 1.03 & 1.8$^{+0.5}_{-0.5}$ & 0.036 $\pm$ 0.004\\%&\\
   L22$^*$ & 161.51763 & 59.08039 & 32.7 $\pm$ 3.6 & 1.06 & 23.7 $\pm$ 3.9 & 1.01 & $<$ 14.1 & 1.00 & 2.0$^{+0.6}_{-0.6}$ & 0.166 $\pm$ 0.010\\%&1.4$\pm0.4^b$\\
   L23 & 161.67129 & 58.89050 & $<$ 11.1 & 0.97 & $<$ 10.8 & 0.92 & $<$ 13.2 & 0.72 & 1.7$^{+0.5}_{-0.5}$ & 0.032 $\pm$ 0.005\\%&\\
   L24$^*$ & 161.25279 & 59.12600 & 11.0 $\pm$ 3.6 & 0.92 & $<$ 12.0 & 0.76 & 17.6 $\pm$ 4.6 & 0.79 & 1.8$^{+0.6}_{-0.5}$ & 0.068 $\pm$ 0.009\\%&3.2$\pm0.4^b$\\
   L25 & 161.41679 & 59.06333 & $<$ 12.0 & 1.25 & $<$ 12.6 & 1.66 & $<$ 15.0 & 6.59 & 1.7$^{+0.6}_{-0.6}$ & 0.024 $\pm$ 0.007\\%&\\
   L27$^*$ & 161.76000 & 58.85086 & 17.4 $\pm$ 3.5 & 0.91 & 19.5 $\pm$ 3.6 & 0.71 & $<$ 13.8 & 1.16 & 2.0$^{+0.8}_{-0.9}$ & 0.078 $\pm$ 0.014\\%&\\
   L28 & 161.58708 & 58.90944 & $<$ 10.5 & 0.74 & $<$ 11.1 & 1.25 & $<$ 13.5 & 1.04 & 1.5$^{+0.5}_{-0.5}$ & 0.025 $\pm$ 0.006\\%&\\
   L29$^*$ & 161.48129 & 59.15453 & 140.5 $\pm$ 3.9 & 2.47 & 64.4 $\pm$ 3.9 & 0.83 & 29.1 $\pm$ 4.8 & 1.46 & 1.8$^{+0.8}_{-0.8}$ & 0.307 $\pm$ 0.043\\%&0.004$^d$\\
   L30$^*$ & 161.29329 & 59.06864 & $<$ 11.7 & 1.02 & 13.3 $\pm$ 4.1 & 1.28 & $<$ 15.0 & 1.59 & 1.5$^{+0.6}_{-0.6}$ & 0.074 $\pm$ 0.007\\%&\\
   L31$^*$ & 161.60379 & 58.89644 & $<$ 10.5 & 1.04 & 20.8 $\pm$ 3.7 & 1.08 & 13.1 $\pm$ 4.3 & 0.89 & 1.5$^{+0.6}_{-0.5}$ & 0.043 $\pm$ 0.004\\%&2.9$\pm0.3^b$\\
   L32$^*$ & 161.41579 & 58.90692 & 18.6 $\pm$ 3.8 & 1.07 & 21.7 $\pm$ 3.9 & 0.98 & 17.6 $\pm$ 5.2 & 1.08 & 1.7$^{+0.8}_{-0.7}$ & 0.047 $\pm$ 0.004\\%&1.3$\pm0.2^b$\\
   L33$^*$ & 161.39608 & 58.84714 & 28.3 $\pm$ 3.7 & 1.19 & 32.5 $\pm$ 3.8 & 0.96 & 26.5 $\pm$ 5.1 & 1.13 & 2.7$^{+1.2}_{-1.2}$ & 0.071 $\pm$ 0.010\\%&\\
   L34$^*$ & 161.22346 & 58.97819 & $<$ 10.8 & 0.94 & 13.3 $\pm$ 3.8 & 0.91 & $<$ 14.1 & 0.74 & 1.5$^{+0.7}_{-0.7}$ & 0.041 $\pm$ 0.008\\%&\\
   L35$^*$ & 161.82546 & 58.92383 & 19.7 $\pm$ 3.7 & 0.95 & 24.4 $\pm$ 3.7 & 0.97 & 18.2 $\pm$ 4.5 & 1.14 & 1.5$^{+0.7}_{-0.7}$ & 0.060 $\pm$ 0.006\\%&\\%3.7$\pm1.3^b$\\
   L36 & 161.53375 & 59.12889 & $<$ 11.7 & 1.40 & $<$ 11.7 & 0.98 & $<$ 14.1 & 1.13 & 1.7$^{+1.1}_{-0.9}$ & 0.016 $\pm$ 0.003\\%&1.7$\pm0.2^b$\\
   L37$^*$ & 161.54575 & 58.87914 & 52.0 $\pm$ 3.6 & 1.01 & 39.9 $\pm$ 3.8 & 1.09 & 16.8 $\pm$ 4.5 & 0.95 & 1.5$^{+0.8}_{-0.7}$ & 0.160 $\pm$ 0.011\\%&\\
   L38 & 161.18704 & 59.13836 & $<$ 12.0 & 0.86 & $<$ 11.7 & 0.81 & $<$ 14.4 & 4.57 & 1.5$^{+0.9}_{-0.9}$ & 0.068 $\pm$ 0.016\\%&\\
   L39$^*$ & 161.55012 & 59.04261 & $<$ 14.1 & 1.12 & $<$ 11.4 & 1.22 & 23.3 $\pm$ 4.5 & 0.81 & 1.4$^{+0.7}_{-0.6}$ & 0.034 $\pm$ 0.003\\%&\\
   L40$^*$ & 161.74304 & 59.10931 & 20.4 $\pm$ 3.7 & 1.09 & 12.2 $\pm$ 4.0 & 0.89 & $<$ 14.1 & 1.05 & 1.5$^{+1.2}_{-1.1}$ & 0.057 $\pm$ 0.005\\%&\\
   L41$^*$ & 161.50129 & 58.91794 & 12.1 $\pm$ 3.7 & 0.86 & 11.6 $\pm$ 3.8 & 0.87 & $<$ 14.1 & 1.19 & 1.4$^{+0.7}_{-0.6}$ & 0.072 $\pm$ 0.012\\%&\\

\hline
\multicolumn{8}{l}{$^*$ detected in at least 1 SPIRE band}\\

\end{tabular}
\end{scriptsize}
\end{table*}

In GOODS-N, 17 of the 24 1.16-mm sources with good identifications are detected in at least 1 SPIRE band, while in Lockman North 30 of the 39 sources with identifications are detected in at least 1 SPIRE band. These figures give a marginally lower detection rate in GOODS-N (71 per cent) as compared to Lockman North (77 per cent), although this is consistent with the slightly deeper nature of the GOODS-N mm-imaging (See \S\ref{sec:radioids}).

As GOODS-N and Lockman Hole North are well-established surveys fields a large number of spectroscopic, and good quality photometric, redshifts are available. In GOODS-N nine sources have spectroscopic redshifts, either from targeted observations of SMGs and radio sources by Chapman et al.\ (2005, and in prep) or from the many magnitude limited redshift surveys of the field which have been conducted (see Barger et al.\ 2008 for a recent compilation of spectroscopic redshifts in GOODS-N). In the Lockman Hole North spectroscopic redshifts are available for four sources, mostly from mid-IR spectroscopy of 24$\,\mu$m selected sources (Fiolet et al.\ 2010), or WIYN spectroscopy of radio sources (Owen \& Morrison 2009). Photometric redshifts are available for an additional nine mm-sources in Lockman Hole North from the catalogue of Strazzullo et al.\ (2010). Here we only consider photometric redshifts which have a quality flag of ``AA'', i.e. the estimated photo-$z$ error is less than $0.2(1+z)$, from Strazzullo et al.\ (2010). In total good quality redshifts are available for 22 of our parent sample of 63 mm-sources. These redshifts are quoted in Tables~\ref{tab:gnspirephot} and~\ref{tab:lhnspirephot}. All of the sources with known spectroscopic and/or reliable optical/near-IR photometric redshifts are detected by SPIRE. This is not particularly surprising, as those sources with redshift information from the optical/near-IR wavelength range tend to be either more intrinsically luminous and/or at slightly lower redshift (Wardlow et al.\ 2011).

To estimate the reliability of our detections, we test random
positions in the SPIRE maps. Test positions are required to be
$>$3\,arcsec from other sources in our input list, and within the
region covered by our 24-$\mu$m and 1.4-GHz catalogue. This is the
typical minimum separation of sources in our input list, due to the
resolutions at 24\,$\mu$m and 1.4\,GHz. If we impose the same
selection criteria on the fluxes recovered for these random positions, 2.7
per cent are detected in any SPIRE band at a significance of
$>$3\,$\sigma$

Taking
this as a guide, if our parent sample of 63 mm-sources were completely spurious then
we would expect to recover SPIRE detections in any band for 2 sources.

\subsection{Submm-mm colours}
\label{sec:smcols}
For the use of submm/mm photometry, either alone or in combination with mid-IR or radio data, to be useful as a redshift estimator there must be a reasonably strong, and unique, matching between submm/mm colour evolution and redshift.

\begin{figure}
\includegraphics[scale=0.25]{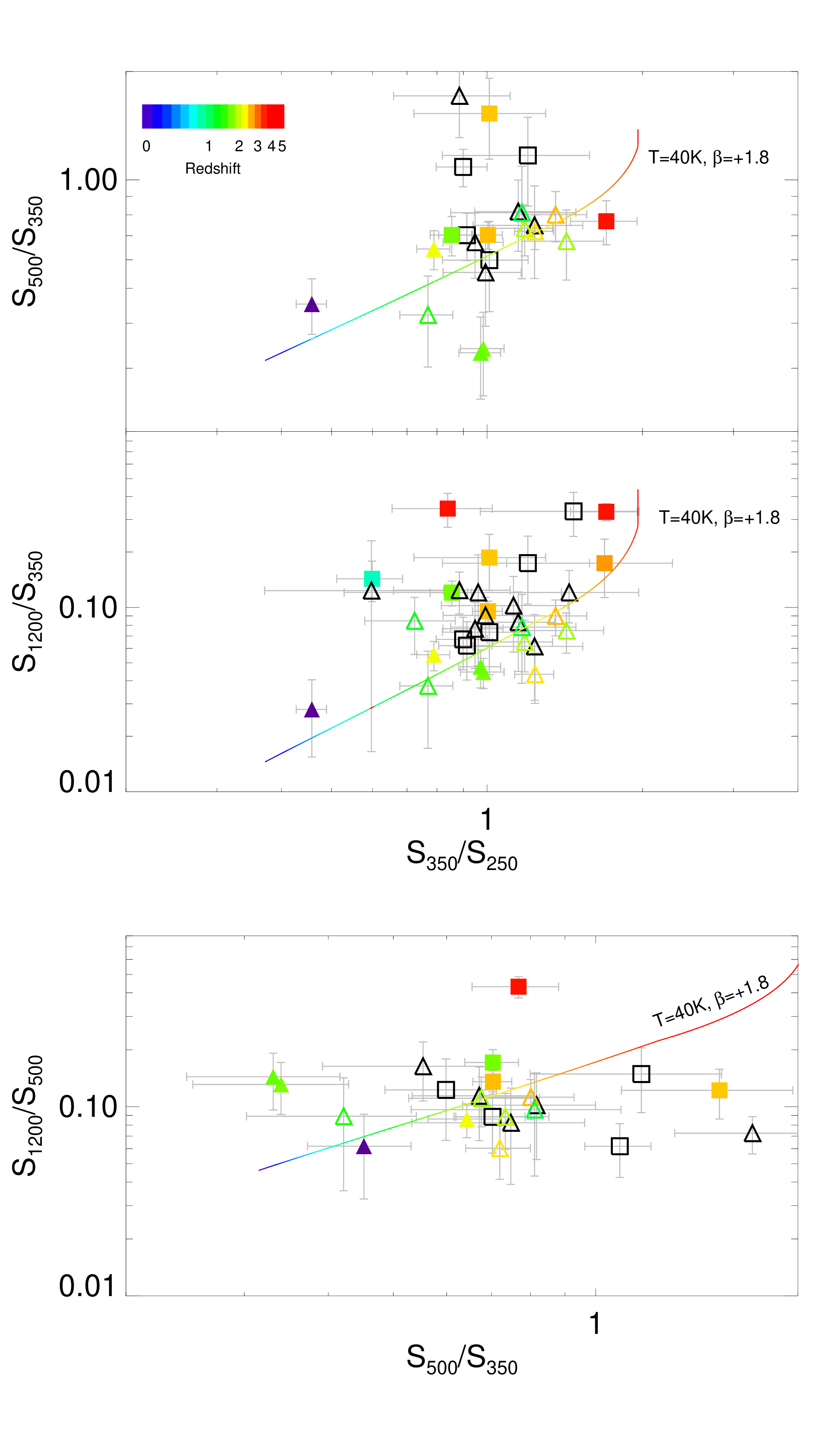}
\caption{Submm-mm colour evolution of mm-selected sources; $S_{500}/S_{350}$ versus $S_{350}/S_{250}$(top) , $S_{\rm 1.2mm}/S_{350}$ versus $S_{350}/S_{250}$ (middle) and $S_{\rm 1.2mm}/S_{500}$ versus $S_{500}/S_{350}$ (bottom). Only sources with $>3\sigma$ detections in each of the relevant SPIRE bands are shown in each panel. The track taken by a modified blackbody, with $T_{\rm d}=40$\,K and optical depth $\tau=(\nu/\nu_0)^{\beta}$ where $\beta=+1.8$ and $\nu_0=c/100\,\mu$m is also shown. Sources located in the GOODS-N field are shown in squares, those in Lockman Hole North are shown as triangles. Symbols are colour-coded by redshift, with filled symbols designating those sources with spectroscopic redshifts and open symbols those with reliable  ($\delta z<0.2(1+z)$) optical/near-IR photometric redshifts. Black open symbols represent sources with no redshift estimate. The submm-mm colours are reasonably well described by our simple modified blackbody model.}
\label{fig:mambocols2}
\end{figure}

In Fig.~\ref{fig:mambocols2}, we compare the SPIRE-mm colours to the expected evolution for a modified blackbody parameterised by a single dust temperature, T$_{\rm D}$, and optical depth $\tau=(\nu/\nu_0)^{beta}$, i.e.
 \[S_{\nu}\propto[1-\exp(-(\frac{\nu}{\nu_0})^{\beta}]B_{\nu}(T_{\rm D}),\]

where $B_{\nu}$ is the Planck function for a single dust temperature $T_{\rm D}$. In order to give a better representation to the mid-IR SED, we replace the blackbody SED on the Wien side of the SED with a power law, $S_{\nu}\propto \nu^{\alpha}$, where $\alpha$ is fixed to -2 (Blain et al.\ 2003; Conley et al.\ 2011). To ensure a smooth SED the power law is used at frequencies where $d\log S/d\log \nu\le\alpha$. As we have a limited number of photometric bands to constrain this model, we fix the dust opacity and optical depth to values of $\beta=+1.8$ and $\nu_0=100\,\mu$m, respectively. We adopt this definition of a modified blackbody throughout this paper. 

The sources with known spectroscopic, or reliable optical/near-IR photometric redshifts from Tables~\ref{tab:gnredshifts} and \ref{tab:lhnredshifts} are coloured coded by redshift.

While the errors on the SPIRE and mm photometry are large, the modified blackbody model can sufficiently explain the observed submm--mm colours of the majority of SPIRE-mm sources. This is qualitatively in agreement with recent results from Kovacs et al.\ 2011, although they warn that more sophisticated models (in particular one which takes account of the range of dust temperatures likely to occur within a star forming galaxy) is needed to match shorter wavelength data (i.e. less than rest frame 30$\,\mu$m). A modified blackbody redshift and temperature are completely degenerate at $z<4$, thus changing the assumed dust temperature will change the predicted submm--mm colour as a function of redshift, but it will not change the position of the model track in colour--colour space. Thus if the dust temperature were known the redshift could be inferred directly, or vice-versa. 

It can be seen that the highest redshift objects are found to have the `reddest' submm/mm colours, i.e. large $S_{\rm mm}/S_{\rm SPIRE}$ or $S_{500}/S_{350}$ flux ratios, supporting claims that identifying red, or `cold', SPIRE/mm sources is an effective way of isolating very high redshift submm galaxies (e.g. Pope et al.\ 2010).

% do I want to estimate beta here?

\begin{table}
\caption{Radio--far IR photometric redshifts, far IR luminosities and dust temperatures for mm-detected sources in the GOODS-N field. Spectroscopic redshifts from the literature are quoted where available. Far IR luminosities (8--1000$\,\mu$m) and dust temperatures are presented for sources with good redshift information. }
\label{tab:gnredshifts}
\begin{tabular}{lllll}
\hline 
\multicolumn{5}{l}{GOODS-N}\\
\hline
ID & Redshift & Radio/Far IR photo-$z$ & L$_{\rm IR}$(L$_{\odot}$) & T$_{\rm D}$(K) \\
\hline\hline
G1$^{*}$  & 4.05$^{a}$ & $^{\dagger}$5.00$^{+0.00}_{-0.4}$ & 13.2$\pm0.1$ & 39$\pm1$\\
G2$^{*}$  & 4.05$^{b}$ & $^{\dagger}$4.20$^{+0.08}_{-0.21}$ & 13.1$\pm0.2$& 43$\pm2$\\
G3$^{*}$  & 2.98$^{c}$ & $^{\dagger}$2.63$^{+0.04}_{-0.06}$ & 13.2$\pm0.1$& 50$\pm2$\\
G4$^{*}$  & 1.996$^{d}$ &$^{\dagger}$2.13$^{+0.00}_{-0.17}$ & 12.7$\pm0.2$& 36$\pm2$\\
G5  &  & $^{\ddagger}$4.5 & & \\
G6$^{*}$  &  & $^{\dagger}$4.07$^{+0.32}_{-0.36}$ & & \\%39 \\
G7  &  & $^{\ddagger}$4.6 & & \\
G8$^{*}$  &   & $^{\wedge}$3.7$^{+0.4}_{-0.2}$ & & \\%39\\
G9  & 0.95$^{e}$ &  $^{\dagger}$0.91$^{+0.13}_{-0.15}$ & & \\
G11$^{*}$  & 2.098$^{c}$ & $^{\dagger}$0.92$^{+0.22}_{-0.41}$ & 12.8$\pm1.3$& 73$\pm36$\\
G13  & & $^{\ddagger}$3.95 & &\\% 35\\
G14  & & $^{\star}$4.1$\pm0.5$ & & \\
G16 & & $^{\ddagger}$3.7 & & \\
G18$^{*}$ & 2.58$^{d}$& $^{\dagger}$2.56$^{+0.18}_{-0.27}$ & 12.5$\pm0.6$& 35$\pm8$\\
G19$^{*}$ & 2.91$^{d}$ &$^{\dagger}$2.91$^{+0.24}_{-0.23}$ & 12.6$\pm0.3$& 36$\pm3$\\
G21$^{*}$ &3.19$^{d}$ & $^{\dagger}$2.82$^{+0.34}_{-0.27}$ & 12.6$\pm0.4$& 38$\pm4$\\
G24 & & $^{\ddagger}$3.1 & & \\
G26$^{*}$ & &  $^{\dagger}$0.24$^{+0.05}_{-0.09}$ & & \\%21\\
G27$^{*}$ & & $^{\dagger}$2.43$^{+0.04}_{-0.18}$ & & \\%39\\
G29$^{*}$ & & $^{\dagger}$2.81$^{+0.26}_{-0.31}$ & & \\%38\\
G30$^{*}$ & & $^{\dagger}$2.41$^{+0.42}_{-0.26}$ & & \\%34\\
G31$^{*}$ & & $^{\dagger}$0.25$^{+0.03}_{-0.11}$ & & \\%37\\
G34 & &$^{\ddagger}$2.2 & & \\
G41$^{*}$  & & $^{\dagger}$1.87$^{+0.17}_{-0.21}$ & & \\%34\\
\hline

\multicolumn{5}{l}{$^*$ detected in at least 1 SPIRE band}\\
\multicolumn{5}{p{8cm}}{$^{\dagger}$ Redshift estimate from simultaneous fitting of L$_{\rm IR}$-T$_{\rm D}$ and L$_{\rm IR}$-L$_{1.4 GHz}$ relations}\\
\multicolumn{5}{p{8cm}}{$^{\ddagger}$ Redshift estimate from mm-1.4\,GHz spectral slope}\\
\multicolumn{5}{p{8cm}}{$^{\wedge}$ Redshift estimate from fitting of L$_{\rm IR}$-T$_{\rm D}$ as radio flux boosted by AGN activity.}\\
\multicolumn{5}{p{8cm}}{$^{\star}$ Radio-Far IR photo-$z$ estimate from Dunlop et al.\ (2004)}\\
\multicolumn{5}{l}{$^{a}$ Spectroscopic redshift from Daddi et al.\ (2009a)}\\
\multicolumn{5}{l}{$^{b}$ Spectroscopic redshift from Daddi et al.\ (2009b)}\\
\multicolumn{5}{l}{$^{c}$ Spectroscopic redshift from Chapman et al., in prep.}\\
\multicolumn{5}{l}{$^{d}$ Spectroscopic redshift from Chapman et al.\ (2005)}\\
\multicolumn{5}{l}{$^{e}$ Spectroscopic redshift from Barger et al.\ (2008)}\\
\end{tabular}
\end{table}
\begin{table}
\caption{Radio--far IR photometric redshifts, far IR luminosities and dust temperatures for mm-detected sources in the Lockman Hole North field. Spectroscopic, or reliable ($\delta z<0.2(1+z)$) optical/near-IR photometric, redshifts from the literature are quoted where available. Far IR luminosities (8--1000$\,\mu$m) and dust temperatures are presented for sources with good redshift information.}
\label{tab:lhnredshifts}

\begin{tabular}{lllll}
\hline 
\multicolumn{5}{l}{Lockman Hole North}\\
\hline
ID & Redshift & Radio/Far IR photo-$z$ & L$_{\rm IR}$(L$_{\odot}$) & T$_{\rm D}$(K) \\
\hline\hline
L1$^{*}$ & 2.562$^a$ & $^{\dagger}$2.06$^{+0.09}_{-0.18}$ & 13.3$\pm0.2$& 54$\pm3$\\
L2$^{*}$ &  & $^{\dagger}$4.09$^{+0.27}_{-0.28}$ & & \\
L3 &  & $^{\ddagger}$1.8 & & \\
L4 &  & $^{\ddagger}$4.4 & & \\
L5$^{*}$ & 3$\pm0.42^b$ & $^{\dagger}$3.30$^{+0.07}_{-0.29}$ & 13.1$\pm0.2$&43$\pm2$ \\
L6$^{*}$ & 2.03$^{c}$ & $^{\dagger}$1.81$^{+0.06}_{-0.17}$ & 12.9$\pm0.2$&43$\pm2$ \\
L7$^{*}$ &  & $^{\dagger}$2.74$^{+0.11}_{-0.20}$ & & \\
L8$^{*}$ &  & $^{\dagger}$2.07$^{+0.20}_{-0.23}$ & & \\
L9 &  & $^{\ddagger}$3.9 & & \\
L10$^{*}$ &  & $^{\dagger}$3.01$^{+0.28}_{-0.21}$ & & \\
L11$^{*}$ & 1.95$^{c}$ & $^{\dagger}$1.71$^{+0.18}_{-0.13}$ & 12.9$\pm0.2$& 41$\pm3$\\
L12$^{*}$ & 2.2$\pm0.2^b$ & $^{\dagger}$2.64$^{+0.08}_{-0.23}$ &12.7$\pm0.2$ & 33$\pm2$\\
L13$^{*}$ &  & $^{\dagger}$0.32$^{+0.18}_{-0.21}$ & & \\
L14$^{*}$ & 2.3$\pm0.5^b$ & $^{\dagger}$2.81$^{+0.16}_{-0.18}$ &12.9$\pm0.2$ & 37$\pm2$ \\
L15$^{*}$ & 2.8$\pm0.2^b$ & $^{\dagger}$3.16$^{+0.09}_{-0.16}$ & 13.3$\pm0.2$& 45$\pm2$ \\
L16$^{*}$ &  & $^{\dagger}$3.24$^{+0.51}_{-0.33}$ & & \\
L17$^{*}$ &  & $^{\dagger}$2.65$^{+0.29}_{-0.19}$ & & \\
L18$^{*}$ &  & $^{\dagger}$2.15$^{+0.11}_{-0.24}$ & & \\
L19 &  & $^{\ddagger}$4.1 & & \\
L21$^{*}$ &  & $^{\dagger}$2.52$^{+0.31}_{-0.20}$ & & \\
L22$^{*}$ & 1.4$\pm0.4^b$ & $^{\dagger}$0.38$^{+0.14}_{-0.28}$ & 12.3$\pm0.5$& 32$\pm4$\\
L23 &  &$^{\ddagger}$3.6  & & \\
L24$^{*}$ & 3.2$\pm0.4^b$ & $^{\dagger}$2.52$^{+0.43}_{-0.20}$ &12.7$\pm0.5$ & 40$\pm5$\\
L25 &  & $^{\ddagger}$3.5 & & \\
L27$^{*}$ &  & $^{\dagger}$1.62$^{+0.45}_{-0.27}$ & & \\
L28 &  & $^{\ddagger}$3.8 & & \\
L29$^{*}$ & 0.004$^d$ & $^{\dagger}$0.38$^{+0.28}_{-0.33}$ &8.0$\pm0.3$ &22$\pm2$ \\
L30$^{*}$ &  & $^{\dagger}$0.71$^{+0.20}_{-0.41}$ & & \\
L31$^{*}$ & 2.9$\pm0.3^b$ & $^{\dagger}$2.44$^{+0.21}_{-0.24}$ & 12.7$\pm0.4$& 40$\pm4$\\
L32$^{*}$ & 1.3$\pm0.2^b$ & $^{\dagger}$2.21$^{+0.25}_{-0.27}$ & 12.0$\pm0.4$& 24$\pm2$\\
L33$^{*}$ &  & $^{\dagger}$2.63$^{+0.16}_{-0.34}$ & & \\
L34$^{*}$ &  & $^{\dagger}$2.46$^{+0.50}_{-0.32}$ & & \\
L35$^{*}$ &  & $^{\dagger}$2.14$^{+0.18}_{-0.24}$ & & \\
L36 &  & $^{\ddagger}$4.5 & & \\
L37$^{*}$ & 1.7$\pm0.2^b$ & $^{\dagger}$1.17$^{+0.18}_{-0.14}$ & & \\
L38 &  & $^{\ddagger}$3.6 & & \\
L39$^{*}$ &  & $^{\dagger}$2.59$^{+0.31}_{-0.21}$ & & \\
L40$^{*}$ &  & $^{\dagger}$0.78$^{+0.13}_{-0.46}$ & & \\
L41$^{*}$ &  & $^{\dagger}$1.49$^{+1.01}_{-0.87}$ & & \\

\hline
\multicolumn{5}{l}{$^*$ detected in at least 1 SPIRE band}\\
\multicolumn{5}{p{8cm}}{$\dagger$ Redshift estimate from simultaneous fitting of L$_{\rm IR}$-T$_{\rm D}$ and L$_{\rm IR}$-L$_{1.4 GHz}$ relations}\\
\multicolumn{5}{p{8cm}}{$\ddagger$ Redshift estimate from mm-radio spectral slope as presented by Linder et al.\ (2011)}\\
\multicolumn{5}{l}{ $^{a}$ Spectroscopic redshift from Polletta et al.\ (2006)}\\
\multicolumn{5}{l}{$^{b}$ Photometric redshift from Strazzullo et al.\ (2010)}\\
\multicolumn{5}{l}{$^{c}$ Spectroscopic redshift from Fiolet et al.\ (2010)}\\
%\multicolumn{9}{l}{$^{d}$ Spectroscopic redshift from Owen et al., in prep}\\
\multicolumn{5}{l}{$^{d}$ Spectroscopic redshift from Owen \& Morrison (2009)}\\
\end{tabular}
\end{table}
\subsection{IR Luminosity to Dust Temperature Relation for SPIRE-mm sources}
\label{sec:detwithid}
As a significant fraction of SPIRE-mm sources have good redshift information (22 of 46 SPIRE detected mm-sources) we can investigate the IR luminosity to dust temperature relation directly for our sample of SPIRE-mm sources.

For each source with a spectroscopic or optical/near-IR photometric redshift (excluding the very low-$z$ source L29) we fit a modified blackbody to the SPIRE and MAMBO photometry (and AzTEC and SCUBA where available), assuming a dust emissivity of $\beta=+1.8$, the redshift and the functional form for a modified blackbody given in Section \ref{sec:smcols}. In the Lockman Hole North field PACS 100 and/or 160$\,\mu$m photometry is available from the HerMES survey for six sources and is included in the SED fits. In GOODS-N SCUBA 850$\,\mu$m photometry is available for five sources from the catalogue of Pope et al.\ (2005) and is included in the SED fits. IR luminosities are calculated for each source by integrating the modified blackbody fit in the range 8--1000$\,\mu$m. IR luminosities and dust temperatures for sources with good redshift information in GOODS-N and Lockman Hole North are presented in Tables \ref{tab:gnredshifts} and \ref{tab:lhnredshifts}, respectively.

\begin{figure}
\includegraphics[scale=0.42]{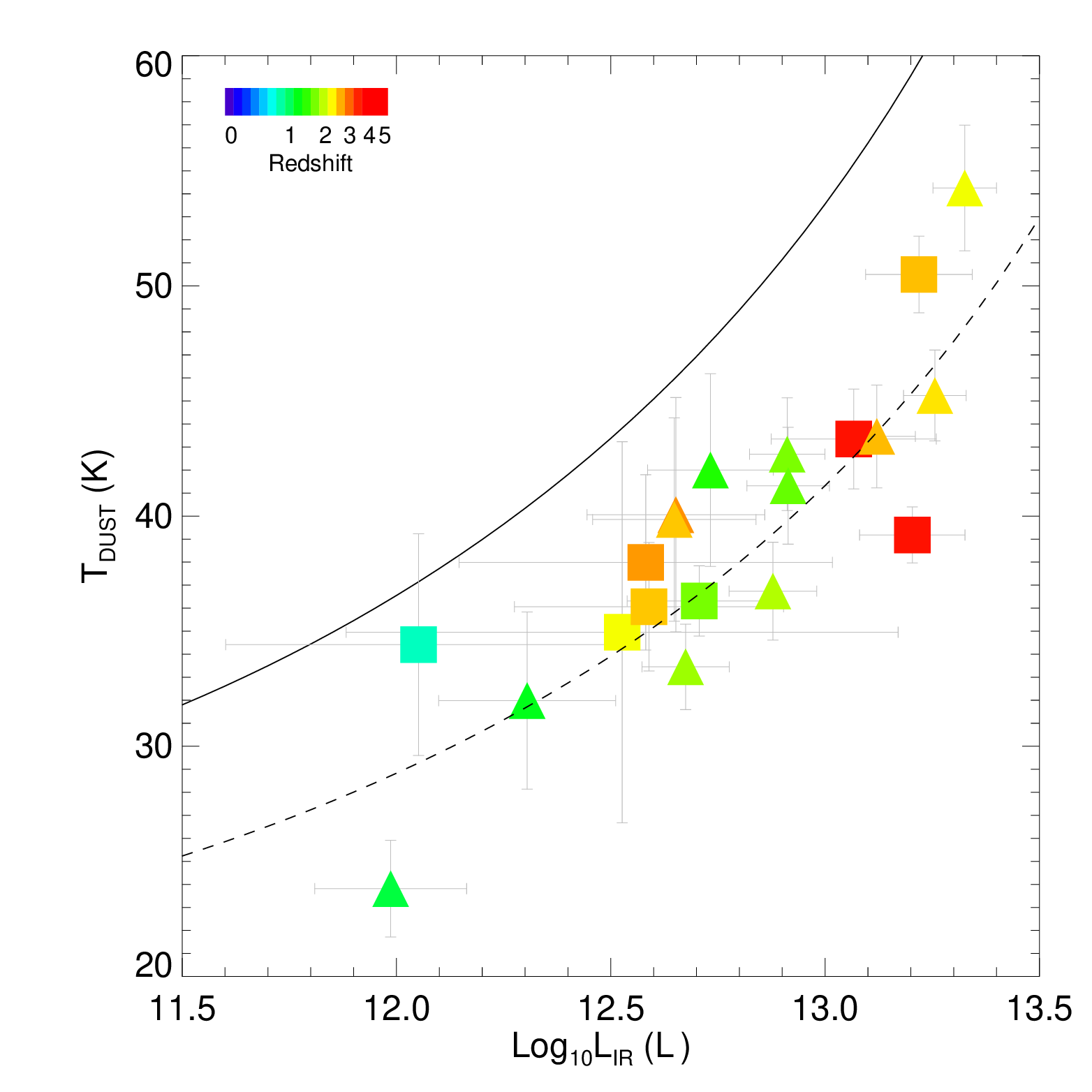}
\caption{IR luminosity vs. dust temperature for those SPIRE-mm sources with good redshift information. Source G11 is omitted as a modified blackbody model is not well-constrained by the observed data. Symbols are colour coded by redshift, with GOODS-N sources denoted by squares and Lockman Hole North sources denoted by triangles. The solid line is the local L$_{\rm IR}$--T$_{\rm D}$ relation from the 1.2\,Jy {\it IRAS} sample as determined by Chapman et al.\ (2003), where we have converted from $S_{100}/S_{60}$ to T$_{\rm D}$ assuming a modified blackbody with optical depth $\tau=(\nu/\nu_0)^{\beta}$ with $\beta=+1.8$ and $\nu_0=c/100\,\mu$m. The dashed line shows a modification to this relation given by fitting Equation \ref{eqn:c03} directly to our SPIRE-mm sources. The best fit parameters are; $\gamma=0.193$,~$\delta=-0.03$ and $C_{\star}=0.3$. L$_{\star}$ is held fixed at $5\times10^{10}$.}

\label{fig:lirtd}
\end{figure}

Fig. \ref{fig:lirtd} shows the best fit modified blackbody IR luminosity vs. dust temperature for our sample of SPIRE-mm sources. One source, G11, is omitted as our modified blackbody model is not well-constrained by the observed data (See Table~\ref{tab:gnredshifts}). Also shown is the local relationship for IRAS 1.2\,Jy sources from Chapman et al.\ (2003); after converting from IRAS  $S_{100}/S_{60}$ colour to T$_{\rm D}$ via our definition of a modified blackbody. It is clear that the SPIRE-mm sources form a tight sequence in L$_{IR}$--T$_{\rm D}$ space. However we can see that our sample is systematically biased towards colder dust temperatures when compared to the local L$_{IR}$--T$_{\rm D}$ relationship (Chapman et al.\ 2003). This bias toward colder dust temperatures for submm/mm selected samples is now well established (Magdis et al.\ 2010; Chapman et al.\ 2010; Magnelli et al.\ 2010).

%Fig. \ref{fig:seds} shows the IR--submm-radio SEDS for all mm-detected sources with good radio identifications in GOODS-N and Lockman North. Where available, 24$\,\mu$m photometry from {\it Spitzer} and PACS 100$\,\mu$m and 160$\,\mu$m photometry from {\it Herschel} are also shown. It can be seen that the Pope et al.\ (2008) SED template does a reasonable job, in most cases, of fitting the full SED of our SPIRE-MAMBO SMGs. 

The bias towards colder dust temperatures causes the L$_{\rm IR}$--T$_{\rm D}$ relation to be much tighter than for an unbiased sample. To parametrise the observed L$_{\rm IR}$--T$_{\rm D}$ relation for mm-selected sources we fit the functional form given by Chapman et al.\ (2003) to our mm-selected sample. Specifically Chapman et al.\ (2003) describe the mean {\it IRAS} colour via,
\begin{equation}
\label{eqn:c03}
\log C_{0}=\log C_{\star}+\delta \log (1+ \frac{L_{star}}{L_{TIR}})+\gamma\log(1+\frac{L_{TIR}}{L_{star}}),
\end{equation}

where $C_{\star}=0.45$, $\delta=-0.02$, $L_{\star}=5\times10^{10}$L$_{\odot}$, and $\gamma=0.16$, while $L_{TIR}$ is the integrated IR luminosity from 3--1100$\,\mu$m and $C_{0}$ is the ratio of the 60-to-100$\,\mu$m {\it IRAS} flux density. We fit Eqn. \ref{eqn:c03} to our observed L$_{IR}$--T$_{\rm D}$ relation of our SPIRE-mm sources, adopting a conversion between {\it IRAS} 60-to-100$\,\mu$m flux ratio and T$_{\rm D}$ given by a modified blackbody with $\beta=+1.8$. We leave $L_{\star}$ fixed at the Chapman et al.\ (2003) value as we do not have enough low luminosity sources to constrain this parameter. Our best fit is given by $C_{\star}=0.3$, $\delta=-0.03$ and $\gamma=0.193$. The track taken by our best fit L$_{IR}$--T$_{\rm D}$ relation is shown in Fig. \ref{fig:lirtd}.

% say something about radio excess sources, X-ray?

\section{Discussion}
\label{sec:discussion}
\subsection{Submm photometric redshifts}\label{sec:smphotz}
As shown in \S\ref{sec:smcols} there is a reasonable correspondence between submm/mm colour and redshift, which is likely a by product of our SPIRE-mm sources demonstrating a strong relationship between dust temperature and luminosity, as seen in Fig.~\ref{fig:lirtd}. Thus, if we can assume a relationship between dust temperature and IR luminosity, we can break the temperature--redshift degeneracy and produce photometric redshifts which rely on submm/mm data alone. 

We explore the potential of such an approach here by making use of the 22 SPIRE-mm sources with good redshift information as a test set. We assume that the L$_{\rm IR}$-T$_{\rm D}$ relationship for SPIRE-mm sources is given by our best fit of Eqn.~\ref{eqn:c03} as seen in Fig.~\ref{fig:lirtd}. Additional photometry from PACS and SCUBA is not considered in the photometric redshift estimation as these are only available for a small subset of the sample; we want a fair assessment of the ability of SPIRE and mm-wavelength data alone to estimate the redshift.

\begin{figure}
\includegraphics[scale=0.45]{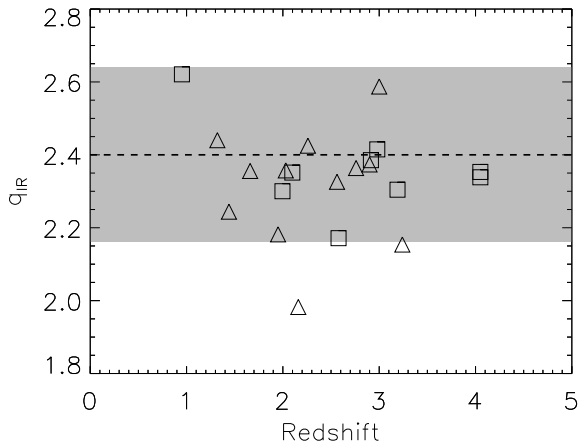}
\caption{Radio-to-Far IR correlation as a function of redshift as probed by the $q_{\rm IR}$ parameter, where $q_{\rm IR}=\log_{10}[($L$_{IR}/3.75\times10^{12}\,{\rm W}) / ($L$_{1.4 GHz}/{\rm \,W\, Hz}^{-1} ) ]$. The dashed line and shaded region represent the median and $\pm1\sigma$ values for SPIRE sources from Ivison et al.\ (2010). Squares and triangles represent sources with reliable redshift information in GOODS-N and Lockman Hole North, respectively. Good agreement is seen with the Ivison et al.\ (2010) values. No significant evolution with redshift is observed, suggesting our single $q_{\rm IR}$ model should introduce no redshift dependant bias to the photometric redshift estimates.}
\label{fig:radfarir}
\end{figure}

\begin{figure*}
\includegraphics[scale=0.48]{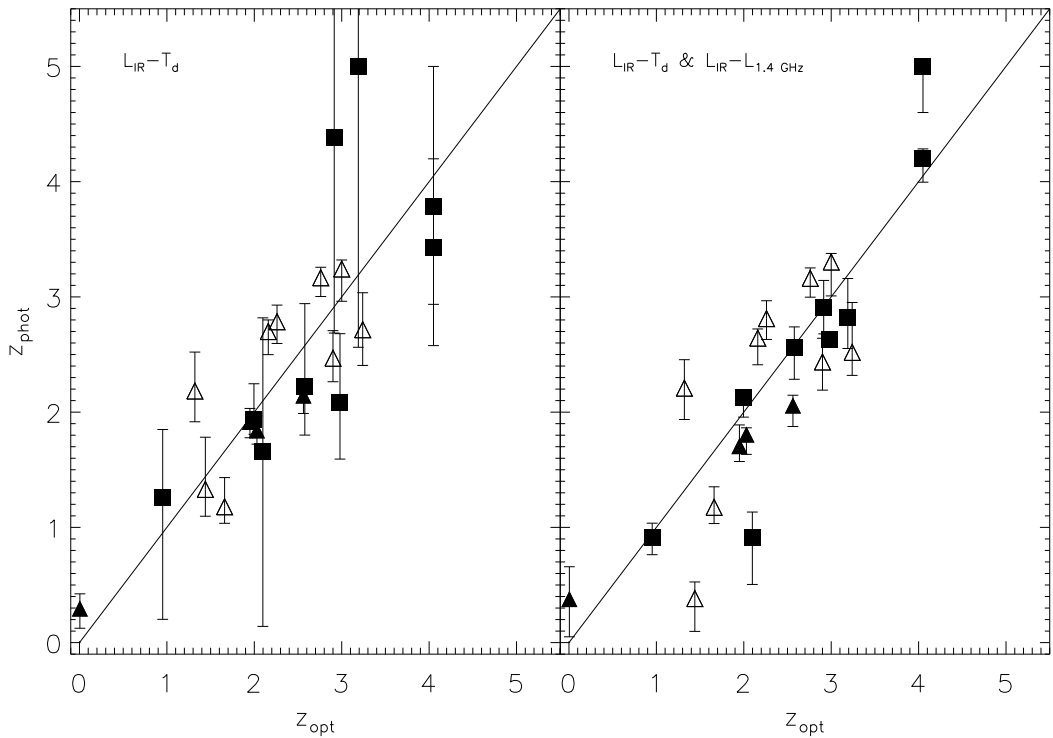}
\caption{Spectroscopic vs photometric redshift estimates for submm/mm photometry alone, assuming a universal L$_{\rm IR}$-T$_{\rm D}$ relation, (left) and combined with radio constraints, assuming a universal radio-to-far IR correlation (right). Filled symbols designate sources with spectroscopic redshifts, while open symbols designate those sources with optical/near-IR photometric redshifts only. For submm/mm photometric redshifts we achieve a mean photo-$z$ accuracy of $|\Delta z|/(1+z)=0.16 \,(<|\Delta z|>=0.51)$, while if radio constraints are included this is reduced to $|\Delta z|/(1+z)=0.15 \,(<|\Delta z|>=0.45)$. Error bars are calculated by finding the redshift range which contains 68 per cent of the probability. }
\label{fig:smphotz}
\end{figure*}

Fig. \ref{fig:smphotz} compares our photometric redshift estimates to the known redshifts for these sources. The mean photometric redshift accuracy is found to be $|\Delta z|/(1+z)=0.16$ with the mean offset $(<|\Delta z|>=0.51)$. All of our SPIRE-mm sources are detected in the radio at 1.4\,GHz so we can hope to improve on our photometric redshift estimates by also requiring that the sources lie on the radio-to-Far IR correlation (van der Kruit 1971). As our SPIRE photometry is close to the peak of the Far-IR emission at $z\sim1-3$, it is not sufficient to assume a spectral index between the SPIRE/mm bands and 1.4\,GHz as is typically done for submm-radio photo-$z$ (i.e. Carilli \& Yun 1999). Here we assume the ratio of L$_{\rm IR}$ to L$_{1.4 GHz}$ is given by a constant value $q_{\rm IR}=\log_{10}[($L$_{IR}/3.75\times10^{12}\,{\rm W}) / ($L$_{1.4 GHz}/{\rm \,W\, Hz}^{-1} ) ]=2.4$ (Ivison et al.\ 2010). We also assume that the radio SED is adequately described by a power law, $S_{\nu}\propto\nu^{\alpha}$, with $\alpha=-0.75$ (Ibar et al.\ 2010). Fig.~\ref{fig:radfarir} shows $q$ as a function of redshift for the 22 SPIRE-mm sources with good redshifts. Good agreement is seen between our sample and the median (and standard deviation) of $q$ for SPIRE sources as measured by Ivison et al.\ (2010). Importantly no evolution with redshift is observed, similar to Ivison et al.\ (2010), and hence our single $q$ model should introduce no redshift dependant bias. As none of the sources in Fig.~\ref{fig:radfarir} lie significantly off the radio-to-far IR correlation, we are confident that potential contributions to the radio flux density from AGN activity may be ignored. 

To get the best estimate of the redshift we simultaneously fit the L$_{\rm IR}$-T$_{\rm D}$ and L$_{\rm IR}$-L$_{1.4 GHz}$ relations, assuming the Far-IR emission is described by a modified blackbody. After including the radio-to-Far IR correlation the photometric redshift accuracy improves to $|\Delta z|/(1+z)=0.15$ or $(<|\Delta z|>=0.45)$. If we consider only those sources with spectroscopic redshifts the accuracy improves again to $|\Delta z|/(1+z)=0.12$ or $(<|\Delta z|>=0.35)$. Interestingly this same improvement is not seen for the L$_{\rm IR}$-T$_{\rm D}$ only method when the sample is restricted to only those with spectroscopic redshifts ($|\Delta z|/(1+z)=0.16$ or $(<|\Delta z|>=0.55)$).

While finding the redshift that gives the maximum likelihood redshift (i.e. the best fit) is the primary aim, it is also of interest to examine the probability distribution to see how wide a range in redshift also gives a reasonable fit. For each source we calculate the redshift range in which 68 per cent of the probability is enclosed. These points are shown as error bars in Fig. \ref{fig:smphotz}. Using this approach, the typical estimated redshift error for photometric redshifts using only the L$_{\rm IR}$-T$_{\rm D}$ is $\sigma_{z}=0.6$, while including the radio-to-Far IR correlation reduces this to $\sigma_{z}=0.4$. Encouragingly these numbers are quite similar to the true error measured by comparing to known redshifts.

Many previous works have attempted to use a combination of radio and and ground-based submm/mm data to estimate photometric redshifts. Aretxaga et al.\ (2007) estimated submm-radio photometric redshifts for 120 850-$\mu$m sources detected in the SHADES survey (Mortier et al.\ 2005), 58 of which had previously known spectroscopic or robust optical/near-IR photometric redshifts. They make use of two distinct photo-$z$ techniques; a simple 850$\,\mu$m to 1.4\,GHz spectral index estimator, similar to that proposed by Carilli \& Yun (1999), and full fitting of the Far-IR to radio SED to a set of well-defined templates. Comparing the results of these two techniques to the known redshifts gives a mean photometric redshift error of $|\Delta z|=0.9$ for the spectral index method, and $|\Delta z|=0.6$ for the template fitting method.

More recently Biggs et al.\ (2011) used a similar spectral index approach to estimate photometric redshifts for 870-$\mu$m sources detected in the LESS survey of ECDFS. Comparing the 48 sources which have spectroscopic redshifts, or reliable 17-band photometric redshifts (Wardlow et al.\ 2011), the photometric redshift accuracy is found to be  $|\Delta z|/(1+z)=0.25$ or a mean deviation of $<|\Delta z|>=0.68$.

Thus it is clear that photometric redshifts estimated using the L$_{\rm IR}$-T$_{\rm D}$ relation alone offer a slight improvement over radio-submm redshifts, while using a combination of the L$_{\rm IR}$--T$_{\rm D}$ relation, and radio-far IR correlation, the accuracy of submm/mm-radio photometric redshifts can be increased by almost a factor of 2. By contrast, a similar study by Dannerbauer et al.\ (2010) using PACS 100 and 160$\,\mu$m, in combination with mid-IR, radio, and submm/mm (but not SPIRE), data find that the accuracy of photometric redshifts using IR templates is significantly worse if PACS data is included. This highlights both the need for SPIRE data, which is at or close to the peak wavelength of the far IR emission at these redshifts, as well as the need for improved SED models in the mid IR (5--30$\,\mu$m.) 

% add something something about EMU

\subsection{Redshift distribution of submm/mm sources}\label{sec:smzdist}
The redshift distribution of sources selected at different submm and mm wavelengths has been a subject of much debate. Many previous works have claimed that sources selected at mm wavelengths are preferentially found at higher redshifts than those at $\sim850\,\mu$m (Dannerbauer et al.\ 2004; Younger et al.\ 2007,  Greve et al.\ 2008, Coppin et al.\ 2009). In contrast, others have claimed that the $\sim$850$\,\mu$m and mm-selected populations are at similar redshifts, with the apparent differences originating from selection effects (Wardlow et al.\ 2011). 

Thus it is of interest to re-examine this using our sample of SPIRE-mm sources. While only 22 of the 46 SPIRE detected mm-sources have reliable spectroscopic or photometric redshifts, as we have seen in \S\ref{sec:smphotz}, it is possible to estimate the redshifts of SPIRE-mm sources with reasonable accuracy using a combination of submm/mm and radio data alone.

As the submm photometric redshift errors are large, and our sample of SPIRE-mm sources small, we prefer to use the full probability distribution function (PDF) for each source rather than simply taking the maximum likelihood estimate of the photometric redshift fit in constructing the redshift distribution. This has the advantage of giving a more robust reconstruction of the redshift distribution (Sheth \& Rossi 2010), although it should be remembered that the distribution produced in this way is essentially a convolution of the true redshift distribution and the typical photo-$z$ error.

In addition to the SPIRE-mm sources we can also estimate the redshifts of the 16 SPIRE undetected mm-sources, using the well-established radio-mm spectral index (Carilli \& Yun 1999). For the sources in Lockman Hole North Lindner et al.\ (2011) quote estimates of the redshift for all 39 1.2-mm sources with radio identifications, using, where possible, a combination of 1.2-mm, 1.4\,GHz and lower frequency GMRT 325 and 610\,MHz observations. We use their redshift estimates for the nine sources without SPIRE counterparts in the Lockman Hole North field. In GOODS-N we again use the spectral index approach as described by Carilli \& Yun (1999) to estimate redshifts for the seven SPIRE undetected mm-sources, assuming the same values for $\alpha_{radio}$ and $\alpha_{submm}$ as Lindner et al.\ (2011), $\alpha_{radio}=-0.68$ and $\alpha_{submm}=3.2$,  to ensure consistency between the two samples. The one exception to this is G14, otherwise known as HDF850.1, for which we used the radio-Far IR photometric redshift of $z=4.1\pm0.5$ determined by Dunlop et al.\ 2004. To assess how reliable the mm-radio spectral index technique is for this sample we compare redshift estimates produced via this method to the known redshifts for the 22 sources in our sample with good redshift information. We find a photo-$z$ accuracy of $|\Delta z|/(1+z)=0.24$, or a mean difference $<|\Delta z|>=0.6$, for photometric redshifts estimated from the radio-mm spectral index alone. 

The full set of SPIRE--mm--radio and mm--radio photometric redshifts is presented in Tables \ref{tab:gnredshifts} and \ref{tab:lhnredshifts} for the GOODS-N and Lockman Hole North fields, respectively.

\begin{figure*}
\includegraphics[scale=0.28]{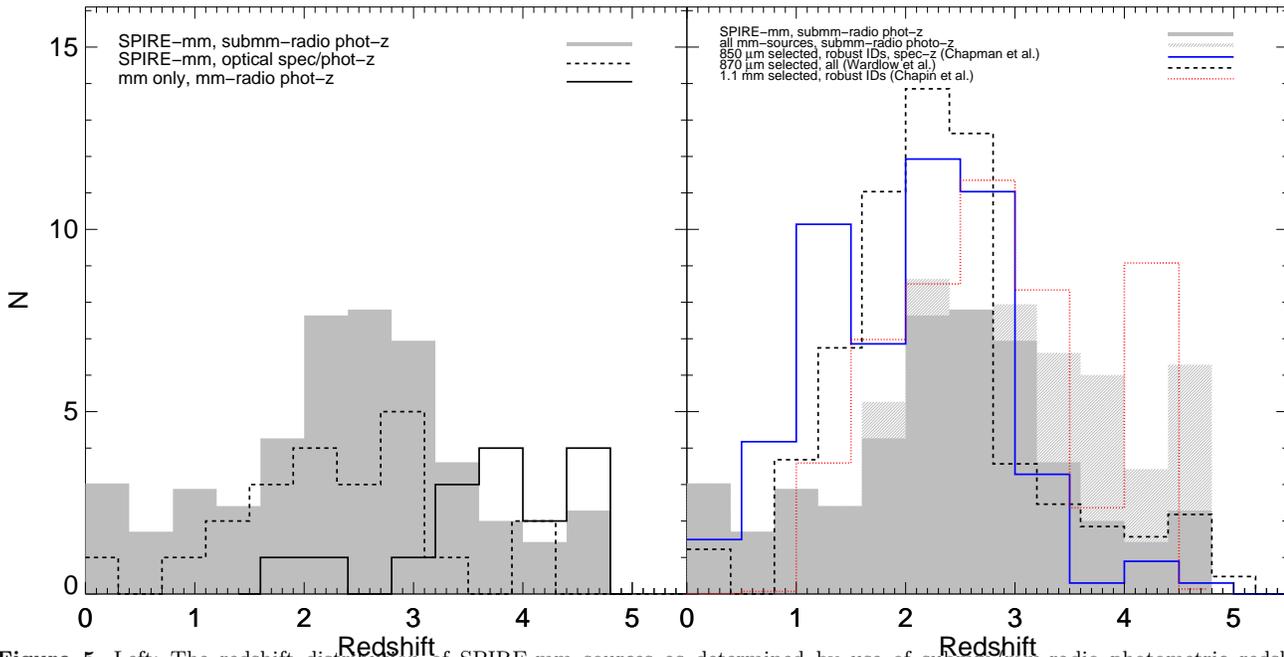}

\caption{Left: The redshift distribution of SPIRE-mm sources as determined by use of submm/mm-radio photometric redshifts. SPIRE-detected mm-sources are shown as a grey shaded histogram. The distribution of SPIRE-mm sources with
 spectroscopic redshifts or reliable optical/near-IR photometric redshifts from Tables ~\ref{tab:gnspirephot} and \ref{tab:lhnspirephot} is shown by the dashed line. Redshift estimates from the radio-mm spectral index for sources from our parent mm-sample without SPIRE detections are shown as a solid line. Right: The redshift distribution of SPIRE-mm sources (grey shaded) and all mm-sources (grey hatched) compared to 850$\,\mu$m selected sources (solid blue line) from a compilation by Chapman et al.\ (in prep.), 1.1-mm selected sources (dotted red line) from Chapin et al.\ (2009.) and 870$\,\mu$m sources (dashed line) from Wardlow et al.\ (2011). In each case the histogram is scaled to the total number of sources in our sample. The mm-selected sources are seen to peak at higher redshifts than those selected at $\sim850\,\mu$m, with a significant excess at $z>3$ made up of primarily mm-only (i.e. SPIRE undetected) sources. }
\label{fig:nz}
 \end{figure*}

The redshift distribution for our SPIRE-mm sources and SPIRE undetected mm-sources is shown in Fig.~\ref{fig:nz}.  SPIRE-mm sources show a strong peak at $z\sim2$--3, similar to other known submm/mm samples. Interestingly, the redshift distribution for SPIRE undetected sources shows that they are predominately found at $z>3$. Only four of the SPIRE detected sample are estimated to be at $z>3.5$, while 13 of the 16 SPIRE undetected mm-sources are found above this redshift. This is not surprising as we have already found in \S\ref{sec:smcols} that the sources with the largest mm-to-SPIRE colours are found to be at the highest redshifts.

However this large population of SPIRE undetected mm-sources at high redshift is somewhat at odds with the redshift distributions of previous submm surveys, which are not limited by SPIRE. Shown for comparison is the redshift distribution of 850$\,\mu$m selected sources, from a compilation by Chapman et al.\ (in prep.), 1.1-mm sources in GOODS-N from Chapin et al.\ (2009; henceforth C09), and 870$\,\mu$m sources in the LABOCA Extended {\it Chandra} Deep Field South Submm Survey (LESS; Wei\ss\ et al.\ 2009) from Wardlow et al.\ (2011; henceforth W11). Comparing the redshift distribution of SPIRE-mm sources and the submm (850/870$\,\mu$m) selected samples of Chapman et al.\ (2005) and Wardlow et al.\ (2011; W11) we find a clear deficit of $z<2.5$ SPIRE-mm sources. In addition, once the mm-only sources are included we find many more $z>3.5$ sources than either Chapman et al.\ or W11. Much better agreement is seen with the C09 sample, although this work shows somewhat fewer sources at low-$z$ (i.e. $z<1$), potentially due to the smaller area probed in GOODS-N compared to our larger sample. A K-S test suggests that the C09 redshifts and our mm-selected sample have a $>99$ per cent chance of being drawn from the same underlying redshift distribution.

One potential reason for the disagreement between the redshift distribution of the mm and 850/870$\,\mu$m-sources is the depth of radio data available; in this work we have used the two deepest blank field images of the sky at 1.4\,GHz available, with a typical rms noise of $\sim3$-4$\,\mu$Jy. This is critical as of the 13 $z>3.5$ SPIRE undetected mm-sources, nine are fainter than 30$\,\mu$Jy at 1.4\,GHz. C09 also make use of the Morrison et al.\ (2010) radio data in GOODS-N. In ECDFS the 1.4\,GHz imaging has an rms of 6.5$\,\mu$Jy at its deepest, and thus the W11 will miss the counterparts for many of the highest redshift sources. Similarly the Chapman et al.\ list is based on lower quality radio data over several fields, and thus is also unlikely to identify the highest redshift submm sources.

W11 attempt to account for submm sources which do not have robust optical/near-IR or radio counterparts by investigating the redshift distribution of optical-near IR sources neighbouring their submm detected sample, and this correction is included in the redshift distribution shown in Fig. \ref{fig:nz}. However while this means that the W11 redshift distribution is not limited by the depth of the radio (or mid-IR) imaging used for identification, this correction is limited by the depth of the available optical/near-IR imaging. W11 estimate that even after this correction, 20 per cent of LESS sources are still unaccounted for. 

\begin{figure*}
\includegraphics[scale=0.4]{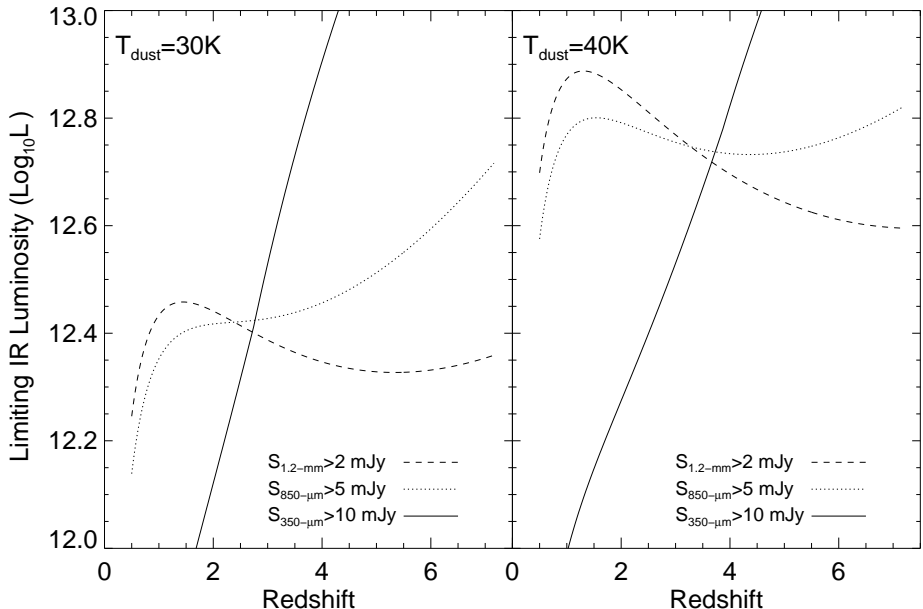}
\caption{Limiting IR luminosity as a function of redshift for flux limited selection at 3 submm/mm-wavelengths; $S_{350}>10$\,mJy (solid line), $S_{850}>5$\,mJy (dotted line) and $S_{1.2mm}>2$\,mJy (dashed line). In each case we have assumed a modified blackbody SED with dust temperature T$_{\rm D}=30\,$K (left) and T$_{\rm D}=40\,$K (right). In both panels, 850$\,\mu$m is more sensitive in the redshift range $1<z<3$, while 1.2-mm is more sensitive at $z>3$.}
\label{fig:mmsel}
\end{figure*}

While it appears that our work here confirms that radio-detected, mm-selected samples appear at higher redshift than those selected at $\sim850\,\mu$m, it is difficult to determine whether this result is not simply a product of our use of deeper 1.4\,GHz radio data for identifications. To investigate the possibility that the $K$-correction between submm (i.e. 850$\,\mu$m) and mm-wavelengths (i.e. 1.2-mm) is the cause for the difference in the redshift distribution in Fig.~\ref{fig:mmsel} we plot the limiting IR luminosity assuming a single modified blackbody with dust temperature 30K and 40K as a function of redshift. The flux limits at 350$\,\mu$m, 850$\,\mu$m and 1.2\,mm are taken to be 10, 5, and 2\,mJy, respectively. The 350$\,\mu$m and 1.2-mm limits were chosen to be representative of the sample under study here, while the 850$\,\mu$m limit was chosen to be representative of current generation surveys at, or near, this wavelength (Wei\ss\/ et al.\ 2009; Mortier et al.\ 2006). 

From Fig.~\ref{fig:mmsel} it is clear that, irrespective of dust temperature, 850$\,\mu$m is marginally more sensitive in the redshift range $1<z<3$, while 1.2-mm is more sensitive at $z>3$. We can also see that the 350$\,\mu$m limiting luminosity increases rapidly as a function of redshift, crossing the 1.2-mm limit at $z\sim3$. This explains the lack of SPIRE-mm sources seen above this redshift.

While the difference in luminosity is small, the shape of the luminosity function at these extreme luminosities is very steep (Chapman et al.\ 2005; W11). At $1<z<2$ the 850$\,\mu$m and 1.2-mm sensitivities differ by $\sim0.1$ decades in L$_{IR}$. Using the Chapman et al.\ (2005) determination of the IR luminosity function, extrapolating to the limiting luminosity shown in the left panel of Fig.~\ref{fig:mmsel} for a T$_{\rm D}=$30K SED (L$_{IR}=12.4$\,L$_{\odot}$; $\log_{10}\phi\sim-5.3$\,Mpc$^3$\,dec.$^{-1}$) and assuming the full volume of our parent surveys between $1<z<3$ (0.24 deg.$^2$; $5.4\times10^{6}$\,Mpc$^3$) we could miss $\sim5$ 850$\,\mu$m detected sources via a 1.2-mm selection. Similarly in the redshift range $3<z<5$, assuming the sensitivities differ by $\sim0.2$ decades in L$_{IR}$ and again using the Chapman et al.\ (2005) luminosity function estimate at L$_{IR}=12.4$\,L$_{\odot}$, we could detect an extra $\sim10$ sources via a 1.2-mm selection at the quoted flux limits. These numbers are crudely in line with the excess, and decrement, of sources we observe in redshift distribution of our mm-selected sources at $1<z<3$ and $z>3$, respectively. Performing a similar exercise assuming T$_{\rm D}=$40K would give a decrement of $\sim1$ and excess of $\sim2$ sources at $1<z<3$ and $3<z<5$, respectively. Thus if the evolution of the 850$\,\mu$m--1.2\,mm $K$ correction is the cause of the observed differences in the redshift distribution it must be relatively cold (T$_{\rm D}\sim$30K) sources which make up the discrepent population.

Of course these calculations assume a particular optical depth $\tau=(\nu/\nu_0)^{\beta}$, although it is worth noting that the differences in sensitivity between 850$\,\mu$m and 1.2\,mm are enhanced for lower values of $\beta$, and are relatively insensitive to $\nu_0$.

\subsection{Contribution of mm-selected sources to the SSFRd}

Now that we have some estimate of the redshift for all of our parent of radio detected mm-selected sample, we can estimate
their contribution to the IR luminosity density ($\rho_{\rm IR}$) and hence the star formation rate density of the Universe (SFRd). Given the wide redshift distribution of mm-sources we estimate $\rho_{\rm IR}$ in four redshift bins; $0.5<z<1.5$,~$1.5<z<2.5$,~$2.5<z<3.5$ and $3.5<z<5$. The redshift bins were chosen so as to match previous measurements (i.e. C05, W11), while still maintaining a reasonable number of sources per bin. The contribution of each source is estimated by taking the measured IR luminosity ($L_{IR}$) and dividing by the maximum volume a source could be detected in ($V_{\rm max}$). Incompleteness in the mm-sample is taken into account by dividing each object by the completeness at its mm-wavelength flux density. Completeness estimates for the 1.2-mm source list in Lockman North are given by Lindner et al.\ (2011). Unfortunately Penner et al.\ (2011) do not provide completeness estimates for their source list in GOODS-N, so we calculate the completeness as a function of 1.16-mm flux density by injecting sources at a range of flux densities into their combined MAMBO-AzTEC image. Sources are considered recovered if they are found within 3 arcsec of the injected position with a peak signal-to-noise of $\sigma>3.8$.  

For each source we wish to use the most reliable redshift available. For the 22 sources with spectroscopic or good optical/near-IR photometric redshifts we use these measurements, while for the 25 SPIRE-mm without prior redshift estimates we use our submm/mm-radio photo-$z$'s from \S\ref{sec:smphotz}. For the remaining 16 mm-only sources we use estimates from the radio-mm spectral index, as described in \S\ref{sec:smzdist}.

However, given the large photometric redshift errors of the submm/mm-radio and radio-mm spectral index estimates which form the bulk of our sample we estimate $\rho_{\rm IR}$ via a Monte-Carlo approach in which realisations are producing by placing each source at a redshift randomly drawn from the probability distribution function (PDF) of the submm/mm-radio fitting. In each of the realisations sources with good spectroscopic redshifts are held fixed. Sources with optical/near-IR photometric redshifts are assumed to have Gaussian PDFs with the width determined by the quoted errors from Table \ref{tab:lhnspirephot}. Similarly mm-only sources with radio-mm spectral index redshifts are assumed to have Gaussian PDFs with width given by the typical $\Delta z$ found in \S\ref{sec:smzdist}; $\Delta z=0.6$. In addition mm-only sources are assumed to be at a dust temperature of 35K.

In each Monte-Carlo realisation we assess the contribution of each source to each redshift bin by taking the sum $\sum_i^n L_{\rm IR}^i/V_{\rm max}^i$, where sum $n$ runs over the sources which happen to fall in that redshift bin in a given realisation, and \lir\ and $V_{\rm max}$ for source $i$ are calculated using the best fit modified blackbody. To calculate $V_{\rm max}$, we find the minimum and maximum redshift at which an observed source would still be found in our sample, i.e. $>3.8\sigma$ at 1.16 or 1.2-mm, depending on the field, and $>5\sigma$ at 1.4\,GHz, taking into account variations in both the luminosity distance and $K$ correction. The posterior probability of $\rho_{\rm IR}$ and the mean redshift in each bin is then inferred from a histogram of the Monte-Carlo realisations, fitting a Gaussian to return the peak likelihood and variance.

Based on this analysis, our best estimate of the contribution of 
mm-selected sources to the IR luminosity density is given in Table~\ref{tab:lird}.

\begin{table}
\caption{Contribution of mm-selected sources to the IR luminosity density.}
\label{tab:lird}
\begin{tabular}{ccc}
\hline
 Redshift range & $\rho_{\rm IR}$ & $\rho_{\rm SFR}$\\
\hline
& Log$_{10}$~L$_{\odot}$~Mpc$^{-3}$ & Log$_{10}$~M$_{\odot}$~yr$^{-1}$~Mpc$^{-3}$\\

\hline\hline
$0.5<z<1.5$ & 7.1$\pm0.1$ & -2.7$\pm0.1$\\
$1.5<z<2.5$ & 7.72$\pm0.07$ & -2.05$\pm0.07$\\
$2.5<z<3.5$ & 7.86$\pm0.08$ & -1.91$\pm0.06$\\
$3.5<z<5$ & 7.7$\pm0.1$ & -2.1$\pm0.1$\\
\hline
\end{tabular}
\end{table}

\begin{figure}
\includegraphics[scale=0.24]{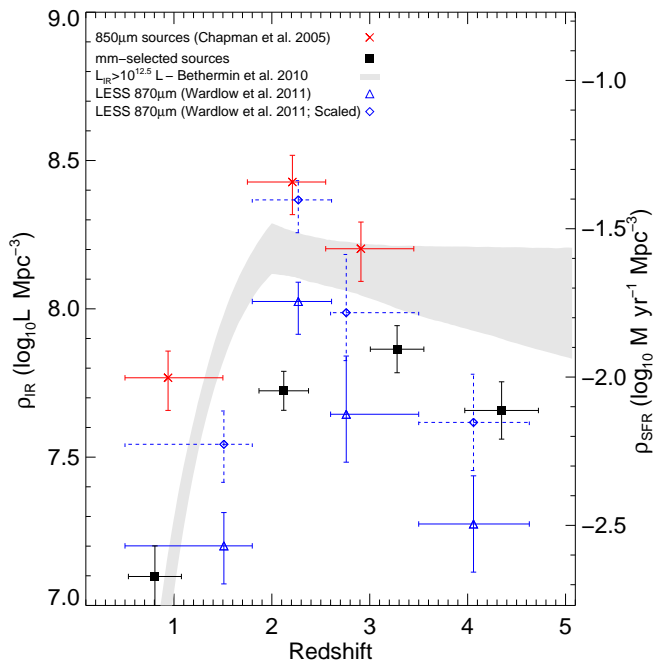}
\caption{Contribution of mm-selected sources to
the IR luminosity density evolution. For comparison,
we show the estimates of Chapman et al.\ (2005) and Wardlow et al.\ (2011) at
$z=0$--4. As ECDFS is underdense at submm wavelengths, Wardlow et al.\ (2011) present two estimates; one based on the observed number of sources in ECDFS, and one which has been scaled to the SHADES 850$\,\mu$m number density (Coppin et al.\ 2006). Also shown is the predicted contribution by $>10^{12.5}$\lsun\ ULIRGs to the IR luminosity density from the empirical model of Bethermin et al.\ (2010). At $z<3$ our estimate is significantly lower than both those found in both previous studies and models, while higher than previous studies at $z\sim4$. These disagreements can be explained when considering the flux limits and biases of the comparison samples.}
\label{fig:lird}
\end{figure}

 Fig.~\ref{fig:lird} compares
these values to previous measures of the contribution to the IR luminosity density by 850$\,\mu$m and LESS 870$\,\mu$m sources (Chapman et al.\ 2005; Wardlow et al.\ 2011), and a prediction of the contribution from $L_{\rm IR}>10^{12.5}\,L_{\odot}$ from the model of Bethermin et al.\ (2010).
At $z\sim2.5$ our estimates appear lower than both the contribution from $\sim850\,\mu$m sources, and the predictions for $L_{\rm IR}>10^{12.5}\,L_{\odot}$ sources. However there are a number of ways to account for this. Firstly, as we have seen in \S\ref{sec:smzdist} mm-selected surveys are less sensitive than their 850$\,\mu$m counterparts at $1<z<3$. However mm-surveys must be less sensitive in general as the areal density of mm-detected sources is much lower than equivalent 850$\,\mu$m samples. Including sources without radio identifications, our parent sample contains 82 mm-detected sources in 0.24 deg.$^2$, while the LESS survey detect 126 sources at 870$\,\mu$m in a 0.25 deg.$^2$ and SHADES detected 120 sources at 850$\,\mu$m across two fields totalling 0.22 deg.$^2$.

Given this, it is puzzling that our estimate at $z\sim4$ is somewhat larger than that found for 870$\,\mu$m sources by W11, however as we have seen in \S\ref{sec:smzdist} there are good reasons to expect a significantly larger contribution at $z>3$, as mm-surveys are more sensitive at these redshifts than their 850$\,\mu$m counterparts, and we have the advantage of significantly deeper radio data with which to identify our mm-sources. The W11 analysis is missing the 45 per cent of LESS SMGs that do not have identifications, similarly Chapman et al.\ (2005) only have radio identifications for $\sim$70 per cent. Here we have radio identifications for 78 per cent of our parent mm-detected sample.

It is of interest that our measurement of the SFRd at $z\sim 4$, while higher than the W11 estimate for 870$\,\mu$m sources, is still somewhat short of that predicted for IR luminous galaxies (Log$_{10}$\,L$_{\rm IR}>12.5$L$_{\odot}$) by Bethermin et al.\ (2010), and $\sim 100$ lower than the total SFRd estimated from a combination of UV, radio and IR measurements (Hopkins \& Beacom 2006). Thus while selecting sources at mm-wavelengths offers an effective way to isolate luminous starbursts at high redshift ($z>3$), it is clear that mm-selected sources, at least at the imaging depth possible with current facilities, do not make a significant contribution to the SFRd at any redshift. 

Due to the strong dust temperature bias, mm-selected sources never contain the bulk of IR luminous sources at any redshift. Taking the Bethermin et al.\ (2010) prediction as a guide, our mm-selected sample never make up more than $\sim50$ per cent of the expected SFRd of IR luminous sources (Log$_{10}$\,L$_{\rm IR}>12.5$). We can see the reason for this clearly in Fig.~\ref{fig:mmsel}, at T$_{\rm D}=30$K we could expect our mm-sample to reach this level, while at T$_{\rm D}=40$K we are only sensitive to IR sources above a luminosity of Log$_{10}$\,L$_{\rm IR}\gs12.7$. Thus for the Bethermin et al.\ (2010) prediction to hold this shortfall must be made up of IR sources with Log$_{10}$\,L$_{\rm IR}\ls12.7$ and T$_{\rm D}\gs30$K. 

\section{Conclusions}\label{sec:conclusion}
We have presented an analysis of 63 radio-detected, mm-selected sources. Our parent sample of mm-sources was drawn from the recent MAMBO 1.2-mm imaging of Lockman Hole North by Lindner et al.\ (2011) and the combined MAMBO-AzTEC 1.16-mm image of GOODS-N presented by Penner et al.\ (2011). Radio identifications are found for the Penner et al.\ (2011) sample, with counterparts for 24 of the 41 1.16-mm sources presented. These new identifications were combined with the 39 (out of 41) radio identifications presented by Lindner et al.\ (2011) for their MAMBO sample to give a sample of 63 radio-detected, mm-sources. We have cross-matched this mm-selected sample with SPIRE data from the HerMES project, making use of a map based technique which reduces the problem of confusion in the SPIRE data by using radio and 24$\,\mu$m positions as a prior. The main results of this work are as follows;

\begin{itemize}
\item After comparing with the SPIRE data we found 47 mm-sources with a detection in one of more SPIRE band, giving a detection rate of 74 per cent.
%\item The SPIRE-mm colours of SPIRE detected mm-sources were found to be qualitatively consistent with a single modified blackbody, with optical depth $\tau=(\nu/\nu_0)^{\beta}$, where dust $\beta=+1.8$ and $\nu_0=100\,\mu$m.
\item For SPIRE detected mm-sources with spectroscopic or good quality optical/near-IR photometric redshifts, we found a tight correlation between dust temperature (T$_{\rm D}$) and IR luminosity (L$_{\rm IR}$). This correlation is found to be offset to colder dust temperatures from the local relationship (e.g. Chapman et al.\ 2003).
\item Using the observed tight relationship between dust temperature (T$_{\rm D}$) and IR luminosity (L$_{\rm IR}$) to break the degeneracy between redshift and dust temperature we have demonstrated that photometric redshifts from submm/mm data alone offer an accuracy of $|\Delta z|/(1+z)=0.16 \,(<|\Delta z|>=0.51)$. If constraints from the radio-far IR correlation are included this is improved to $|\Delta z|/(1+z)=0.15 \,(<|\Delta z|>=0.45)$.
\item Via a combination of spectroscopic, optical/near-IR photometric and submm/mm-radio photometric redshifts, the redshift distribution of mm-selected sources was presented. We found that the redshift distribution of mm-selected sources peaks at higher redshifts than those selected at $\sim850\,\mu$m, with a long tail to high-$z$ ($z>3.5$) made up of primarily mm-only (i.e. SPIRE undetected) sources. This difference may be explained by a combination of the depth of the radio data used to identify the sources, as well as the evolving $K$-correction between submm and mm-wavelengths. Good agreement was found between this study and previous mm-selected samples.
\item We measured the contribution of mm-selected sources to the star formation rate density (SFRd) of the Universe. We found that mm-selected sources do not contribute significantly to the global SFRd, nor do they make up the bulk of IR luminous sources, at any redshift. 
\end{itemize}

\section*{Acknowledgements}
We thank the anonymous referee for suggestions which greatly enhanced this work.

LW and SJO were supported by the Science and Technology Facilities Council
[ST/F002858/1].\\
JSD acknowledges the support of the Royal Society via a Wolfson Research Merit award, and the support of the European Research Council via the award of an Advanced Grant.\\
AF, GM, LM and MV were supported by the Italian Space Agency (ASI âHerschel Scienceâ Contract I/005/07/0).\\
SPIRE has been developed by a consortium of institutes led by Cardiff
Univ.\ (UK) and including Univ.\ Lethbridge (Canada); NAOC (China);
CEA, LAM (France); IFSI, Univ.\ Padua (Italy); IAC (Spain); Stockholm
Observatory (Sweden); Imperial College London, RAL, UCL-MSSL, UK ATC,
Univ.\ Sussex (UK); Caltech, JPL, NHSC, Univ.\ Colorado (USA). This
development has been supported by national funding agencies: CSA
(Canada); NAOC (China); CEA, CNES, CNRS (France); ASI (Italy); MCINN
(Spain); SNSB (Sweden); STFC (UK); and NASA (USA). \\

The data presented in this paper will be released through the {\em Herschel} Database in Marseille HeDaM ({hedam.oamp.fr/HerMES})\\

\label{lastpage}


\begin{thebibliography}{99}
%\bibliography{lit}
%\bibliographystyle{mn2e}
\bibitem[Akaike(1974)]{1974ITAC...19..716A} Akaike, H.\ 1974, IEEE 
Transactions on Automatic Control, 19, 716 
\bibitem[Alexander et al.(2005)]{2005Natur.434..738A} Alexander, D.~M., 
Smail, I., Bauer, F.~E., Chapman, S.~C., Blain, A.~W., Brandt, W.~N., 
\& Ivison, R.~J.\ 2005, \nat, 434, 738
%\bibitem[Alonso-Herrero et al.(2006)]{2006ApJ...640..167A} Alonso-Herrero, 
%A., et al.\ 2006, \apj, 640, 167
\bibitem[Aretxaga et al.(2003)]{2003MNRAS.342..759A} Aretxaga, I., Hughes, 
D.~H., Chapin, E.~L., Gazta{\~n}aga, E., Dunlop, J.~S., 
\& Ivison, R.~J.\ 2003, \mnras, 342, 759 
\bibitem[Aretxaga et al.(2007)]{2007MNRAS.379.1571A} Aretxaga, I., et al.\ 
2007, \mnras, 379, 1571
%\bibitem[Austermann et al.(2010)]{2010MNRAS.401..160A} Austermann, J.~E., 
%et al.\ 2010, \mnras, 401, 160 
\bibitem[Barger et al.(1998)]{1998Natur.394..248B} Barger, A.~J., Cowie, 
L.~L., Sanders, D.~B., Fulton, E., Taniguchi, Y., Sato, Y., Kawara, K., 
\& Okuda, H.\ 1998, \nat, 394, 248 
\bibitem[Barger et al.(2008)]{2008ApJ...689..687B} Barger, A.~J., Cowie, 
L.~L., \& Wang, W.-H.\ 2008, \apj, 689, 687
%\bibitem[Berta et 
%al.(2007)]{2007A&A...476..151B} Berta, S., et al.\ 2007, \aap, 476, 151
%\bibitem[B{\'e}thermin et al.(2010)]{2010arXiv1003.0833B} B{\'e}thermin, 
%M., Dole, H., Cousin, M., \& Bavouzet, N.\ 2010, arXiv:1003.0833 
\bibitem[B{\'e}thermin et al.(2010)]{2010arXiv1010.1150B} B{\'e}thermin, 
M., Dole, H., Lagache, G., Le Borgne, D., 
\& P{\'e}nin, A.\ 2010, arXiv:1010.1150 
%\bibitem[Biggs 
%\& Ivison(2008)]{2008MNRAS.385..893B} Biggs, A.~D., \& Ivison, R.~J.\ 2008, \mnras, 385, 893 
\bibitem[Biggs et al.(2010)]{2010MNRAS.408..342B} Biggs, A.~D., Younger, 
J.~D., \& Ivison, R.~J.\ 2010, \mnras, 408, 342
\bibitem[Biggs et al.(2011)]{2011MNRAS.413.2314B} Biggs, A.~D., et al.\ 
2011, \mnras, 413, 2314 
\bibitem[Blain 
\& Longair(1993)]{1993MNRAS.265L..21B} Blain, A.~W., \& Longair, M.~S.\ 1993, \mnras, 265, L21 
\bibitem[Blain(1997)]{1997MNRAS.290..553B} Blain, A.~W.\ 1997, \mnras, 290, 
553
\bibitem[Blain(1999)]{1999MNRAS.309..955B} Blain, A.~W.\ 1999, \mnras, 309, 
955 
\bibitem[Blain et al.(2003)]{2003MNRAS.338..733B} Blain, A.~W., Barnard, 
V.~E., \& Chapman, S.~C.\ 2003, \mnras, 338, 733 
%\bibitem[Cantalupo et al.(2010)]{2010ApJS..187..212C} Cantalupo, C.~M., 
%Borrill, J.~D., Jaffe, A.~H., Kisner, T.~S., 
%\& Stompor, R.\ 2010, \apjs, 187, 212 
%\bibitem[Borys et al.(2002)]{2002MNRAS.330L..63B} Borys, C., Chapman, 
%S.~C., Halpern, M., \& Scott, D.\ 2002, \mnras, 330, L63
%\bibitem[Borys et al.(2003)]{2003MNRAS.344..385B} Borys, C., Chapman, S., 
%Halpern, M., \& Scott, D.\ 2003, \mnras, 344, 385
\bibitem[Capak et al.(2008)]{2008ApJ...681L..53C} Capak, P., et al.\ 2008, \apjl, 681, L53 
\bibitem[Carilli \& Yun(1999)]{1999ApJ...513L..13C} Carilli, C.~L., \& Yun, M.~S.\ 1999, \apjl, 513, L13 
\bibitem[Carilli et al.(2010)]{2010ApJ...714.1407C} Carilli, C.~L., Daddi, 
E., Riechers, D., et al.\ 2010, \apj, 714, 1407
\bibitem[Carilli et al.(2011)]{2011ApJ...739L..33C} Carilli, C.~L., Hodge, 
J., Walter, F., et al.\ 2011, \apjl, 739, L33 
\bibitem[Chapin et al.(2009)]{2009MNRAS.398.1793C} Chapin, E.~L., et al.\ 
2009, \mnras, 398, 1793 
\bibitem[Chapin et al.(2011)]{2011MNRAS.411..505C} Chapin, E.~L., et al.\ 
2011, \mnras, 411, 505 


\bibitem[Chapman et al.(2001)]{2001ApJ...548L.147C} Chapman, S.~C., 
Richards, E.~A., Lewis, G.~F., Wilson, G., 
\& Barger, A.~J.\ 2001, \apjl, 548, L147 
\bibitem[Chapman et al.(2003)]{2003ApJ...588..186C} Chapman, S.~C., Helou, 
G., Lewis, G.~F., \& Dale, D.~A.\ 2003, \apj, 588, 186 
\bibitem[Chapman et al.(2005)]{2005ApJ...622..772C} Chapman, S.~C., Blain, 
A.~W., Smail, I., \& Ivison, R.~J.\ 2005, \apj, 622, 772
\bibitem[Chapman et al.(2010)]{2010MNRAS.409L..13C} Chapman, S.~C., et al.\ 
2010, \mnras, 409, L13 
\bibitem[Coppin et al.(2009)]{2009MNRAS.395.1905C} Coppin, K.~E.~K., et 
al.\ 2009, \mnras, 395, 1905 
%\bibitem[Coppin et al.(2010)]{2010MNRAS.407L.103C} Coppin, K.~E.~K., et 
%al.\ 2010, \mnras, 407, L103 
\bibitem[Cowie et al.(2009)]{2009ApJ...697L.122C} Cowie, L.~L., Barger, 
A.~J., Wang, W.-H., \& Williams, J.~P.\ 2009, \apjl, 697, L122 

\bibitem[Daddi et al.(2009a)]{2009ApJ...694.1517D} Daddi, E., et al.\ 2009a, \apj, 694, 1517
\bibitem[Daddi et al.(2009b)]{2009ApJ...695L.176D} Daddi, E., Dannerbauer, 
H., Krips, M., Walter, F., Dickinson, M., Elbaz, D., 
\& Morrison, G.~E.\ 2009b, \apjl, 695, L176


\bibitem[Dannerbauer et al.(2004)]{2004ApJ...606..664D} Dannerbauer, H., 
Lehnert, M.~D., Lutz, D., Tacconi, L., Bertoldi, F., Carilli, C., Genzel, 
R., \& Menten, K.~M.\ 2004, \apj, 606, 664 
\bibitem[Dannerbauer et al.(2010)]{2010ApJ...720L.144D} Dannerbauer, H., 
Daddi, E., Morrison, G.~E., et al.\ 2010, \apjl, 720, L144 
\bibitem[Dey et al.(1999)]{1999ApJ...519..610D} Dey, A., Graham, J.~R., 
Ivison, R.~J., Smail, I., Wright, G.~S., 
\& Liu, M.~C.\ 1999, \apj, 519, 610 

\bibitem[Downes et al.(1986)]{1986MNRAS.218...31D} Downes, A.~J.~B., 
Peacock, J.~A., Savage, A., \& Carrie, D.~R.\ 1986, \mnras, 218, 31 
\bibitem[Dunlop et al.(2004)]{2004MNRAS.350..769D} Dunlop, J.~S., et al.\ 
2004, \mnras, 350, 769
\bibitem[Eales et al.(1999)]{1999ApJ...515..518E} Eales, S., Lilly, S., 
Gear, W., Dunne, L., Bond, J.~R., Hammer, F., Le F{\`e}vre, O., 
\& Crampton, D.\ 1999, \apj, 515, 518 
\bibitem[Eales et al.(2010)]{2010PASP..122..499E} Eales, S., et al.\ 2010, 
\pasp, 122, 499
%\bibitem[Condon(1974)]{1974ApJ...188..279C} Condon, J.~J.\ 1974, \apj, 188, 
%279 
%\bibitem[Coppin et al.(2005)]{2005MNRAS.357.1022C} Coppin, K., Halpern, M., 
%Scott, D., Borys, C., \& Chapman, S.\ 2005, \mnras, 357, 1022
%\bibitem[Devlin et al.(2009)]{2009Natur.458..737D} Devlin, M.~J., et al.\ 
%2009, \nat, 458, 737 
%\bibitem[Dole et 
%al.(2006)]{2006A&A...451..417D} Dole, H., et al.\ 2006, \aap, 451, 417 


%\bibitem[Dye et al.(2009)]{2009ApJ...703..285D} Dye, S., et al.\ 2009, 
%\apj, 703, 285 
\bibitem[Fernandez-Conde et 
al.(2008)]{2008A&A...481..885F} Fernandez-Conde, N., Lagache, G., Puget, J.-L., \& Dole, H.\ 2008, \aap, 481, 885 
%\bibitem[Furusawa et al.(2008)]{2008ApJS..176....1F} Furusawa, H., et al.\ 
%2008, \apjs, 176, 1 
%\bibitem[Farrah et al.(2008)]{2008ApJ...677..957F} Farrah, D., et al.\ 
%2008, \apj, 677, 957
\bibitem[Fiolet et 
al.(2010)]{2010A&A...524A..33F} Fiolet, N., et al.\ 2010, \aap, 524, A33 
\bibitem[Franceschini et 
al.(1991)]{1991A&AS...89..285F} Franceschini, A., Toffolatti, L., Mazzei, P., Danese, L., \& de Zotti, G.\ 1991, \aaps, 89, 285 

%\bibitem[Giavalisco et al.(2004)]{2004ApJ...600L..93G} Giavalisco, M., et 
%al.\ 2004, \apjl, 600, L93 
%\bibitem[Greve et al.(2005)]{2005MNRAS.359.1165G} Greve, T.~R., et al.\ 
%2005, \mnras, 359, 1165 
\bibitem[Glenn et al.(2010)]{2010MNRAS.409..109G} Glenn, J., et al.\ 2010, 
\mnras, 409, 109
\bibitem[Greve et al.(2008)]{2008MNRAS.389.1489G} Greve, T.~R., Pope, A., 
Scott, D., Ivison, R.~J., Borys, C., Conselice, C.~J., 
\& Bertoldi, F.\ 2008, \mnras, 389, 1489 
\bibitem[Griffin et al.(2010)] {GriffinSPIRE} Griffin, M. et al.,\ 2010, \aap,518, 3

%\bibitem[Hogg(2001)]{2001AJ....121.1207H} Hogg, D.~W.\ 2001, \aj, 121, 1207 
\bibitem[Holland et al.(1999)]{1999MNRAS.303..659H} Holland, W.~S., et al.\ 
1999, \mnras, 303, 659 
\bibitem[Hughes et al.(1998)]{1998Natur.394..241H} Hughes, D.~H., et al.\ 
1998, \nat, 394, 241 
\bibitem[Hughes et al.(2002)]{2002MNRAS.335..871H} Hughes, D.~H., et al.\ 
2002, \mnras, 335, 871 
\bibitem[Ibar et al.(2010)]{2010MNRAS.401L..53I} Ibar, E., Ivison, R.~J., 
Best, P.~N., Coppin, K., Pope, A., Smail, I., 
\& Dunlop, J.~S.\ 2010, \mnras, 401, L53 
%\bibitem[Ivison et al.(1998)]{1998MNRAS.298..583I} Ivison, R.~J., Smail, 
%I., Le Borgne, J.-F., Blain, A.~W., Kneib, J.-P., Bezecourt, J., Kerr, 
%T.~H., \& Davies, J.~K.\ 1998, \mnras, 298, 583 
%\bibitem[Ivison et al.(2002)]{2002MNRAS.337....1I} Ivison, R.~J., et al.\ 
%2002, \mnras, 337, 1 
%\bibitem[Ivison et al.(2004)]{2004ApJS..154..124I} Ivison, R.~J., et al.\ 
%2004, \apjs, 154, 124 
%\bibitem[Ivison et al.(2005)]{2005MNRAS.364.1025I} Ivison, R.~J., et al.\ 
%2005, \mnras, 364, 1025
\bibitem[Ivison et al.(2007)]{2007MNRAS.380..199I} Ivison, R.~J., et al.\ 
2007, \mnras, 380, 199 
\bibitem[Ivison et 
al.(2010a)]{2010A&A...518L..31I} Ivison, R.~J., et al.\ 2010a, \aap, 518, L31 
\bibitem[Ivison et al.(2010b)]{2010MNRAS.404..198I} Ivison, R.~J., Smail, 
I., Papadopoulos, P.~P., Wold, I., Richard, J., Swinbank, A.~M., Kneib, 
J.-P., \& Owen, F.~N.\ 2010b, \mnras, 404, 198


%\bibitem[Lonsdale et al.(2003)]{2003PASP..115..897L} Lonsdale, C.~J., et 
%al.\ 2003, \pasp, 115, 897 
%\bibitem[Kneib et al.(2004)]{2004MNRAS.349.1211K} Kneib, J.-P., van der 
%Werf, P.~P., Kraiberg Knudsen, K., Smail, I., Blain, A., Frayer, D., 
%Barnard, V., \& Ivison, R.\ 2004, \mnras, 349, 1211
%\bibitem[Knudsen et al.(2010)]{2010ApJ...709..210K} Knudsen, K.~K., Kneib, 
%J.-P., Richard, J., Petitpas, G., \& Egami, E.\ 2010, \apj, 709, 210 
\bibitem[Kov{\'a}cs et al.(2006)]{2006ApJ...650..592K} Kov{\'a}cs, A., 
Chapman, S.~C., Dowell, C.~D., Blain, A.~W., Ivison, R.~J., Smail, I., 
\& Phillips, T.~G.\ 2006, \apj, 650, 592 
\bibitem[Kov{\'a}cs et al.(2010)]{2010ApJ...717...29K} Kov{\'a}cs, A., 
Omont, A., Beelen, A., et al.\ 2010, \apj, 717, 29
%\bibitem[Kreysa et al.(1998)]{1998SPIE.3357..319K} Kreysa, E., et al.\ 
%1998, \procspie, 3357, 319


%\bibitem[Laird et al.(2010)]{2010MNRAS.401.2763L} Laird, E.~S., Nandra, K., 
%Pope, A., \& Scott, D.\ 2010, \mnras, 401, 2763 

\bibitem[Levenson et al.(2010)]{2010MNRAS.409...83L} Levenson, L., et al.\ 
2010, \mnras, 409, 83
\bibitem[Lindner et al.(2011)]{2011arXiv1106.0344L} Lindner, R.~R., et al.\ 
2011, arXiv:1106.0344 
\bibitem[Lilly et al.(1999)]{1999ApJ...518..641L} Lilly, S.~J., Eales, 
S.~A., Gear, W.~K.~P., Hammer, F., Le F{\`e}vre, O., Crampton, D., Bond, 
J.~R., \& Dunne, L.\ 1999, \apj, 518, 641
%\bibitem[Lonsdale et al.(2009)]{2009ApJ...692..422L} Lonsdale, C.~J., et 
%al.\ 2009, \apj, 692, 422  
\bibitem[Magdis et al.(2010)]{2010MNRAS.409...22M} Magdis, G.~E., et al.\ 
2010, \mnras, 409, 22
%\bibitem[Magnelli et 
%al.(2009)]{2009A&A...496...57M} Magnelli, B., Elbaz, D., Chary, R.~R., Dickinson, M., Le Borgne, D., Frayer, D.~T., \& Willmer, C.~N.~A.\ 2009, \aap, 496, 57 
\bibitem[Magnelli et 
al.(2010)]{2010A&A...518L..28M} Magnelli, B., et al.\ 2010, \aap, 518, L28 
\bibitem[Marsden et al.(2009)]{2009ApJ...707.1729M} Marsden, G., et al.\ 2009, \apj, 707, 1729
%\bibitem[Mortier et al.(2005)]{2005MNRAS.363..563M} Mortier, A.~M.~J., et 
%al.\ 2005, \mnras, 363, 563 
\bibitem[Men{\'e}ndez-Delmestre et al.(2009)]{2009ApJ...699..667M} 
Men{\'e}ndez-Delmestre, K., et al.\ 2009, \apj, 699, 667 
%\bibitem[Mori{\'c} et al.(2010)]{2010ApJ...724..779M} Mori{\'c}, I., 
%Smol{\v c}i{\'c}, V., Kimball, A., Riechers, D.~A., Ivezi{\'c}, {\v Z}., 
%\& Scoville, N.\ 2010, \apj, 724, 779
\bibitem[Morrison et al.(2010)]{2010ApJS..188..178M} Morrison, G.~E., Owen, 
F.~N., Dickinson, M., Ivison, R.~J., \& Ibar, E.\ 2010, \apjs, 188, 178
\bibitem[Mortier et al.(2005)]{2005MNRAS.363..563M} Mortier, A.~M.~J., et 
al.\ 2005, \mnras, 363, 563
\bibitem[Nguyen et 
al.(2010)]{2010A&A...518L...5N} Nguyen, H.~T., et al.\ 2010, \aap, 518, L5 
\bibitem[Oliver et al.(2010)] {oliverSurvey} Oliver, S. et al.,\ 2010, in prep
\bibitem[Owen 
\& Morrison(2008)]{2008AJ....136.1889O} Owen, F.~N., \& Morrison, G.~E.\ 2008, \aj, 136, 1889 
\bibitem[Owen 
\& Morrison(2009)]{2009ApJS..182..625O} Owen, F.~N., \& Morrison, G.~E.\ 2009, \apjs, 182, 625
%\bibitem[Oliver et al.(2010b)] {oliverCounts} Oliver, S. et al.,\ 2010, \aap, in press
\bibitem[Pascale et al.(2009)]{2009ApJ...707.1740P} Pascale, E., et al.\ 
2009, \apj, 707, 1740 
%\bibitem[Patanchon et al.(2008)]{2008ApJ...681..708P} Patanchon, G., et al.\ 2008, \apj, 681, 708 
%\bibitem[Patanchon et al.(2009)]{2009ApJ...707.1750P} Patanchon, G., et 
%al.\ 2009, \apj, 707, 1750
%\bibitem[Park et al.(2010)]{2010ApJ...717.1181P} Park, S.~Q., et al.\ 2010, 
%\apj, 717, 1181
\bibitem[Penner et al.(2010)]{2010MNRAS.tmp.1667P} Penner, K., et al.\ 
2010, \mnras, in press
\bibitem[Perera et al.(2008)]{2008MNRAS.391.1227P} Perera, T.~A., et al.\ 
2008, \mnras, 391, 1227
\bibitem[Pilbratt et al.(2010)]{Herschel} Pilbratt, G., et al.\ 2010, \aap, 518, 1
\bibitem[Polletta et al.(2006)]{2006ApJ...642..673D} Polletta, M.~d.~C., et 
al.\ 2006, \apj, 642, 673 
%\bibitem[Pope et al.(2005)]{2005MNRAS.358..149P} Pope, A., Borys, C., Scott, D., Conselice, C., Dickinson, M., \& Mobasher, B.\ 2005, \mnras, 358, 149
\bibitem[Pope et al.(2006)]{2006MNRAS.370.1185P} Pope, A., et al.\ 2006, \mnras, 370, 1185 
\bibitem[Pope 
\& Chary(2010)]{2010ApJ...715L.171P} Pope, A., \& Chary, R.-R.\ 2010, \apjl, 715, L171
%\bibitem[Pope et al.(2008)]{2008ApJ...675.1171P} Pope, A., et al.\ 2008, 
%\apj, 675, 1171 

\bibitem[Roseboom et al.(2009)]{2009MNRAS.400.1062R} Roseboom, I.~G., Oliver, S., Parkinson, D., \& Vaccari, M.\ 2009, \mnras, 400, 1062
\bibitem[Roseboom et al.(2010)]{2010MNRAS.409...48R} Roseboom, I.~G., et 
al.\ 2010, \mnras, 409, 48 
%\bibitem[Serjeant et al.(2003)]{2003MNRAS.344..887S} Serjeant, S., et al.\ 
%2003, \mnras, 344, 887
\bibitem[Smail et al.(1997)]{1997ApJ...490L...5S} Smail, I., Ivison, R.~J., 
\& Blain, A.~W.\ 1997, \apjl, 490, L5  
\bibitem[Strazzullo et al.(2010)]{2010ApJ...714.1305S} Strazzullo, V., 
Pannella, M., Owen, F.~N., Bender, R., Morrison, G.~E., Wang, W.-H., 
\& Shupe, D.~L.\ 2010, \apj, 714, 1305
%\bibitem[Savage 
%\& Oliver(2007)]{2007ApJ...661.1339S} Savage, R.~S., \& Oliver, S.\ 2007, \apj, 661, 1339 
%\bibitem [Scharz(1978)]{1978AS...6..461S} Schwarz, Gideon E.\ 1978, Annals of Statistics 6 (2): 461–464
%\bibitem[Scheuer \& Ryle(1957)]{1957PCPS...53..764S} Scheuer, P.~A.~G., \& Ryle, M.\ 1957, Proceedings of the Cambridge Philosophical Society, 53, 764 
%\bibitem[Scott et al.(2002)]{2002MNRAS.331..817S} Scott, S.~E., et al.\ 
%2002, \mnras, 331, 817
%\bibitem[Shupe et al.(2008)]{2008AJ....135.1050S} Shupe, D.~L., et al.\ 
%2008, \aj, 135, 1050]
%\bibitem[Stark \& Parker(1995)]{} Stark, P.B., \& Parker, R.L., 1995, Comp. Stat. 10., 129-141
\bibitem[Swinbank et al.(2008)]{2008MNRAS.391..420S} Swinbank, A.~M., et 
al.\ 2008, \mnras, 391, 420 
\bibitem[Swinyard et al.(2010)] {SPIREcalibration} Swinyard, B., et al.\ 2010, \aap, 518, 4
\bibitem[Tacconi et al.(2008)]{2008ApJ...680..246T} Tacconi, L.~J., et al.\ 
2008, \apj, 680, 246 
\bibitem[Tibshirani (1996)]{LASSO} Tibshirani, R. 1996, J. Royal. Statist. Soc B., 58, 1, 267
\bibitem[ter Braak 2010]{NNLSpath} ter Braak, C.J.F., et al.\ 2010, Genetics, 185, 1045
%\bibitem[Wall et al.(2008)]{2008MNRAS.383..435W} Wall, J.~V., Pope, A., 
%\& Scott, D.\ 2008, \mnras, 383, 435 
\bibitem[van der 
Kruit(1971)]{1971A&A....15..110V} van der Kruit, P.~C.\ 1971, \aap, 15, 110 
%\bibitem[Wang et al.(2004)]{2004ApJ...613..655W} Wang, W.-H., Cowie, L.~L., 
%\& Barger, A.~J.\ 2004, \apj, 613, 655
\bibitem[Wang et al.(2011)]{2011ApJ...726L..18W} Wang, W.-H., Cowie, L.~L., 
Barger, A.~J., \& Williams, J.~P.\ 2011, \apjl, 726, L18 
\bibitem[Wardlow et al.(2011)]{2011MNRAS.415.1479W} Wardlow, J.~L., et al.\ 
2011, \mnras, 415, 1479 
\bibitem[Wei{\ss} et al.(2009)]{2009ApJ...707.1201W} Wei{\ss}, A., et al.\ 
2009, \apj, 707, 1201
\bibitem[Younger et al.(2007)]{2007ApJ...671.1531Y} Younger, J.~D., et al.\ 
2007, \apj, 671, 1531
%\bibitem[Younger et al.(2009)]{2009ApJ...704..803Y} Younger, J.~D., et al.\ 
%2009, \apj, 704, 803
%\bibitem[Tegmark(1997)]{1997ApJ...480L..87T} Tegmark, M.\ 1997, \apjl, 480, 
%L87 
%\bibitem[Yun et al.(2008)]{2008MNRAS.389..333Y} Yun, M.~S., et al.\ 2008, 
%\mnras, 389, 333 
\bibitem[Zou 2006]{aLASSO} Zou, H. 2006, J. Amer. Statist. Assoc., 101, 1418
\end{thebibliography}
\end{document}